\documentclass[lettersize,journal]{IEEEtran}
\usepackage{amsmath,amsfonts}
\usepackage{array}
\usepackage[justification=centering]{subcaption}
\usepackage{textcomp}
\usepackage{stfloats}
\usepackage{url}
\usepackage{verbatim}
\usepackage{graphicx}
\usepackage{cite}

\usepackage{times}
\usepackage{epsfig}
\usepackage{caption}
\usepackage{subcaption}
\usepackage{amssymb}
\usepackage{algorithm}
\usepackage{algorithmicx}
\usepackage{algpseudocode}
\usepackage{enumerate}
\usepackage{multirow}
\usepackage{colortbl}
\usepackage{url}
\usepackage{bbding}
\usepackage{microtype}
\usepackage{array}
\usepackage{float}

\usepackage{lipsum}
\usepackage{mathtools}
\usepackage{dsfont}
\usepackage{diagbox}

\usepackage{xspace}

\makeatletter
\DeclareRobustCommand\onedot{\futurelet\@let@token\@onedot}
\def\@onedot{\ifx\@let@token.\else.\null\fi\xspace}
\def\eg{\emph{e.g}\onedot} 
\def\ie{\emph{i.e}\onedot} 
 
\def\etc{\emph{etc}\onedot} 
 
\def\etal{\emph{et al}\onedot}
\makeatother

\begin{document}

\title{Data Acquisition and Preparation for Dual-reference Deep Learning of Image Super-Resolution}

\author{Yanhui~Guo,~\IEEEmembership{}
        Xiaolin~Wu,~\IEEEmembership{Fellow,~IEEE},~and % <-this % stops a space
        Xiao~Shu,~\IEEEmembership{}
\thanks{ 
Manuscript was accepted by IEEE Transactions on Image Processing.
%  April 19, 2021; revised August 16, 2021. 
% This work was supported by the Natural Sciences and Engineering Research Council 
% of Canada (NSERC). (Corresponding author: Xiaolin Wu.)
% \newline 
% \indent Yanhui Guo, Xiaolin Wu and Xiao Shu are with the Department of Electrical and Computer Engineering,
% McMaster University, Hamilton, ON L8G 4K1, Canad (E-mail: guoy143@mcmaster.ca; xwu@ece.mcmaster.ca; shux@mcmaster.ca).
% \newline
% \indent This paper has supplementary downloadable material available at
% http://ieeexplore.ieee.org, provided by the author. 
}}

\markboth{Journal of \LaTeX\ Class Files,~Vol.~14, No.~8, August~2021}%
{Shell \MakeLowercase{\textit{et al.}}: A Sample Article Using IEEEtran.cls for IEEE Journals}


\maketitle

\begin{abstract}
  The performance of deep learning based image super-resolution (SR) methods depend on how accurately the
  paired low and high resolution images for training characterize the sampling process of real cameras. Low and high resolution
  (LR$\sim$HR) image pairs synthesized by degradation models (e.g., bicubic downsampling) deviate from those in reality; thus the synthetically-trained DCNN SR models work disappointingly
  when being applied to real-world images. To address this issue, we propose a novel data acquisition process to shoot a large
  set of LR$\sim$HR image pairs using real cameras. The images are displayed on an ultra-high quality screen and captured
  at different resolutions. The resulting LR$\sim$HR image pairs can be aligned at very high sub-pixel precision by a novel
  spatial-frequency dual-domain registration method, and hence they provide more appropriate training data for the learning
  task of super-resolution. Moreover, the captured HR image and the original digital image offer dual references to strengthen
  supervised learning. Experimental results show that training a super-resolution DCNN by our LR$\sim$HR dataset achieves higher
  image quality than training it by other datasets in the literature. Moreover, the proposed screen-capturing data collection process
  can be automated; it can be carried out for any target camera with ease and low cost, offering a practical way of tailoring the
  training of a DCNN SR model separately to each of the given cameras.
\end{abstract}


\section{Introduction}
\IEEEPARstart{S}{ingle} image super-resolution (SR), the task of increasing the
spatial resolution and details of a given image, has attracted great attention from the computer vision research communities in the last few decades \cite{huang2015single,yang2010image}.
In recent years, the state of the art of SR has been set and reset by various deep convolutional neural network (DCNN) based techniques \cite{dong2015PAMI, ledig2017photo, zhang2018ECCV}. In addition to the significant improvement in SR performance, these DCNN techniques also provide some important
insights on the designs of DCNN architectures \cite{lim2017enhanced, zhang2018image,yang2020learning, kim2016accurate} and loss functions \cite{johnson2016perceptual,ledig2017photo,sajjadi2017enhancenet,Wang_2018_ECCV_Workshops}.
However, for many real-world problems, the efficacy of a machine learning technique relies not only on the design of the technique itself but also, sometimes even more critically, on the quality and representativeness of the training data.

Low resolution is not only caused by covering a large field of view given the sensor array resolution.
Other contributing factors could be out of focus, camera jittering, or sensor noises.  Downsampling HR images by an artificial degradation function does not account for these real-world imperfections and hence it does not generate true LR images.
However, due to the difficulties of obtaining real LR$\sim$HR image pairs, existing SR techniques still resort to synthetically generated low-resolution (LR) images for training.
The overly-simplistic downsampling operators commonly employed by these techniques, such as bicubic downsampling (BD) and Gaussian downsampling (GD), cannot accurately simulate the complex and compounded process of
capturing a low-resolution image.  When operating on real images,
the DCNNs trained on these synthesized datasets generally cannot superresolve high-frequency details to the same level of clarity and sharpness as when operating on synthetic LR images.
This weakness of synthetic training images is well known to practitioners.

It is very difficult to replace synthetic images with real ones in the training of SR networks.  Ideally, the perfectly aligned LR and HR image pair should be obtained by shooting the same scene with the identical camera optics but with two 2D sensor arrays, one of low and the other of high resolutions.  This could only be realized by using a beam splitter and two sensor arrays.  But such a camera is complex, costly, and not readily available.
There have been a few attempts to collect datasets of paired LR and HR images \cite{chen2019camera,cai2019toward,zhang2019zoom}. Their basic idea is to capture each natural scene twice with lenses of different scale factors (focal lengths). Then, they form an image pair by aligning the two images using image registration techniques, possibly patch by patch.
However, this approach is inherently problematic and it can never achieve the perfect alignment as if we could only swap sensor arrays of low and high resolution.  First of all, swapping lenses will inevitably perturb the camera position, causing disparity between the resulting LR and HR images and consequently changes in depth of field and perspective. The alignment of LR and HR images faces the problems of blurring and occlusion, which cannot be corrected by algorithms.
In addition, as each scene is captured twice, a scene has to be absolutely static between the two shots to prevent mismatches of the paired images due to motions. Thus, any moving subjects, such as animals or humans, cannot be included in the datasets.
The lack of alignment precision between LR and HR training images can throw off the deep learning process, hindering the reconstruction of sharp image details.

To circumvent the above difficulties, we propose to acquire SR training data by shooting images displayed on an ultra-high resolution monitor.
An high quality image $\mathcal{Y}_i$ is displayed and then captured twice by a camera, in high resolution via a long focal lens, and also in low resolution via a short focal lens, generating the captured HR image $Y_i$ and LR image $X_i$ in pair.  Shooting an ultra-high resolution monitor to create paired LR and HR images avoid the issues when aligning $Y_i$ and $X_i$ as discussed in the previous paragraph.  Indeed, now the
registration problem can be modeled exactly by homography transform because the imaged scene is a 3D planar surface.  The flatness of the screen eliminates
the problems of occlusion and defocusing due to depth disparity in pairing.
Even there is a small change of the camera position after lens swap, the perspective change can be easily compensated for by homography transform, which is not the case for existing methods \cite{chen2019camera,cai2019toward,zhang2019zoom}.
In addition, as the scene of the captured image pairs is static, absolutely no
object motions exist to cause the difficulties of occlusion and spatially varying alignment parameters.  In summary, shooting screen can achieve a much higher registration accuracy between the captured LR image $X_i$ and the captured HR image $Y_i$; and in turn, between $X_i$ and $\mathcal{Y}_i$.

To push the registration accuracy to the limit, we propose a novel dual-domain
LR$\sim$HR registration method.  In this method specially designed registration markers are displayed on the screen and imaged by the camera.  They are used to align LR$\sim$HR image pairs to sub-pixel precision by the proposed joint spatial-frequency registration method.

\begin{figure}[ht]
   \centering
   \includegraphics[width=0.48\textwidth]{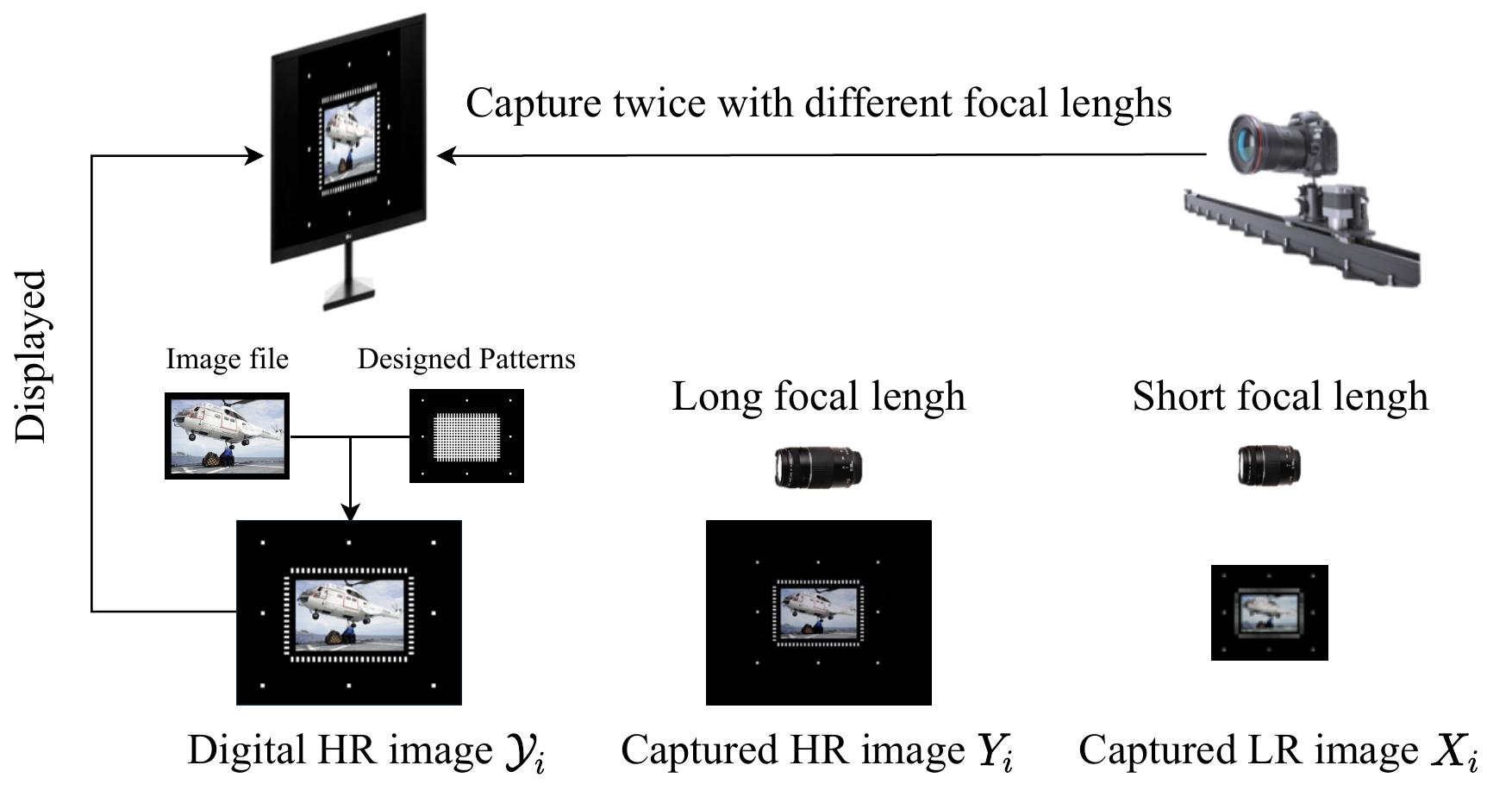}
   \caption{Illustration of the data collection process of DRTI acquisition system.} % (please note the real shooting distance is longer as shown in Table.\ref{table_camera}).}
   \label{Figure_Intro}
\end{figure}

The LR images $X_i$ for training the DCNN SR model are physically acquired rather than algorithmically downsampled from the corresponding HR image $\mathcal{Y}_i$ (digital image file).  Thus, the $(X_i,Y_i)$ image pair provides the deep learning algorithm with the information on pixel PSF, sensor noise statistics, lens characteristics, \etc.
These physical effects are very difficult to model analytically as they are complex and compounded to each other, and are best left to be implicitly modeled by a data-driven DCNN.  But the captured HR screen image $Y_i$ by itself is not an ideal ground truth, because it is not an exact copy of the original $\mathcal{Y}_i$. To compensate for this we also include $\mathcal{Y}_i$ as a part of the ground truth, and
introduce the dual references $(Y_i,\mathcal{Y}_i)$ as a combined ground truth for the supervised learning of the SR task.  The resulting monitor-induced dual-reference training image (DRTI) triplets $(X_i,Y_i,\mathcal{Y}_i)$ are used to train DCNN SR models.  The role of the original HR digital image $\mathcal{Y}_i$ is to provide structural
information on high frequency features that might be compromised by the analog operation of screen shooting.
Fig.~\ref{Figure_Intro} illustrates the proposed DRTI acquisition process.

Except the aforementioned difficulties of the real SR datasets collection, another thorny issue is that the real SR datasets are not camera-agnostic.
This problem is caused by varied specifications (\eg pixel pitch, noise model, sensitivity and \etc) of different camera sensors. These
differences make the statistics of collected LR images dependent on the camera used for shooting.
The DCNN SR model trained with this camera-specific dataset thus inclines to overfit the camera and performs poorly outside the dataset. One simple way
is to collect a large number of image pairs using all cameras for which the SR DCNN is designed. However, this is not practical, because collecting the
diverse large dataset by shooting many scenes using different cameras is laborious, time consuming and costly.
Moreover, the DCNN SR model trained by the mixed dataset will not yield the best results for a specific camera.
The proposed LR$\sim$HR image data collection and preparation process can overcome the above weakness.
Once the shooting environment is set up, the whole image acquisition process, including capturing and alignment, can be easily automated without manual intervention. The data throughput can reach 500 image pairs per hour.
Moreover, the same procedure can be conveniently and quickly repeated for different cameras.  This opens up the possibility to train a DCNN SR model for any target camera for the best possible performance.

The rest of this paper is organized as follows. After a brief
review of related works in Section \ref{related_work},
we present, in Section \ref{DRTI_system}, the
laboratory setup and details of our data collection process.
Section \ref{sec_image_resig} introduces and rationalizes the proposed spatial-frequency dual-domain registration method.
Section \ref{sec_loss_func} presents the dual-reference loss function for training a DCNN model with the DRTI triplets.
Section \ref{results} reports our empirical results on laboratory datasets and real-world datasets.
The experiments demonstrate the advantages of the DRTI datasets.
Section \ref{conclusion} concludes the paper.

\section{Related Work}\label{related_work}

\subsection{Single Image Super-Resolution}

In the past decade single image super-resolution has evolved from a topic of classical image processing \cite{zhangwu2008,sun2008image,yang2010image}
to a showcase of deep learning applications.
The latter class of SR methods aim to learn nonlinear mappings from LR to HR images typically using large paired datasets.
After the pioneering DCNN method SRCNN \cite{dong2015PAMI}
a large number of DCNN models have been proposed to improve SR performance by adopting more crafted neural network architecture designs,
such as residual block \cite{kim2016accurate,ledig2017photo,zhang2018image}, attention mechanism, including channel attention \cite{zhang2018image},
non-local attention \cite{liu2018non,mei2021image}, and adaptive patch aggression \cite{zhou2020cross}.
Very recently, vision Transformer (ViT) \cite{dosovitskiy2020vit}, which was inspired by the success of Transformer in NLP, 
was proposed to better characterize correlations between different parts of an image.  ViT showed great promise on the SR task, as demonstrated by the methods SwinIR \cite{liang2021swinir} and Restormer \cite{zamir2021restormer}.

In addition to works on suitable network architectures, some unsupervised SR methods have been studied, attempting to dispense with paired training data altogether.
Ulyanov \etal \cite{Ulyanov_2018_CVPR} explored the low-level image statistics prior of a generator network and applied it to solving the super-resolution problem.
Shocher \etal \cite{Shocher_2018_CVPR} investigated a zero-shot SR method which exploits the internal recurrence of
information inside a single image, and only needs to train a small image-specific CNN at test time, without any predetermined training pairs.
Extending their work, Gandelsman \etal \cite{Soh_2020_CVPR} presented meta-transfer learning for zero-shot super-resolution, which
adopts transfer learning to seek a good initial point for the zero-shot learning.  However, these unsupervised methods cannot match the performance of the supervised counterparts.

\subsection{Image Super-Resolution Datasets}
There are several popular datasets that have been extensively used for training and testing in SR, such as Set5 \cite{bevilacqua2012low}, Set14 \cite{zeyde2010single}, Urban100 \cite{huang2015single} and DIV2K \cite{timofte2017ntire}.
In all of these datasets, the LR images were generated from the HR images via synthetic degradation like bicubic downsampling or Gaussian blurring followed by direct downsampling \cite{dong2012nonlocally} . Xu \etal \cite{Xu2019TowardsRS} proposed a method to generate approximate realistic
training data by simulating the imaging process of digital cameras. Yoo \etal \cite{Jaejun2020Rethink} proposed a CutBlur method to augment training data for super-resolution. Jeon \etal \cite{jeon2018enhancing} and Wang \etal \cite{wang2019learning} tried to improve super-resolution performance by using stereo low resolution images.
Besides, Generative adversarial networks (GANs) have also been applied to generate realistic degraded images \cite{Manuel2019,yuan2018unsupervised,Adrian2018}. Nevertheless, all these synthetic datasets are still far away from the real images.

Recently, some researchers have attempted to capture real-world image pairs for training DCNN SR models. Qu \etal \cite{Qu2016} obtained paired LR$\sim$HR images by placing a beam splitter in the optical path of two identical camera sensors. But their dataset only includes face images. Köhler \etal \cite{Thomas2020} implemented hardware
binning on the sensors to collect real data, but it only has 14 scenes. Zhang \etal \cite{zhang2019zoom} collected 500 scenes using multiple focal lengths. However, they cannot register the paired images precisely because of
perspective misalignment. Chen \etal \cite{chen2019camera} printed 100 images to postcards and captured LR$\sim$HR image pairs, but the models trained on this dataset cannot generalize well to real-world scenes and they cannot register the image pairs precisely by only using
SIFT\cite{lowe2004distinctive} features. Cai \etal \cite{cai2019toward} collected
595 LR$\sim$HR image pairs using two digital single-lens reflex (DSLR) cameras. However, they meet the perspective misalignment as same as \cite{zhang2019zoom} and they cannot capture paired data from dynamic scenes. Different from them, our DRTI acquisition system
can automatically generate enough precisely aligned LR$\sim$HR training data at a low cost.

\begin{figure*}[ht]
   \centering
   \includegraphics[width=1\textwidth]{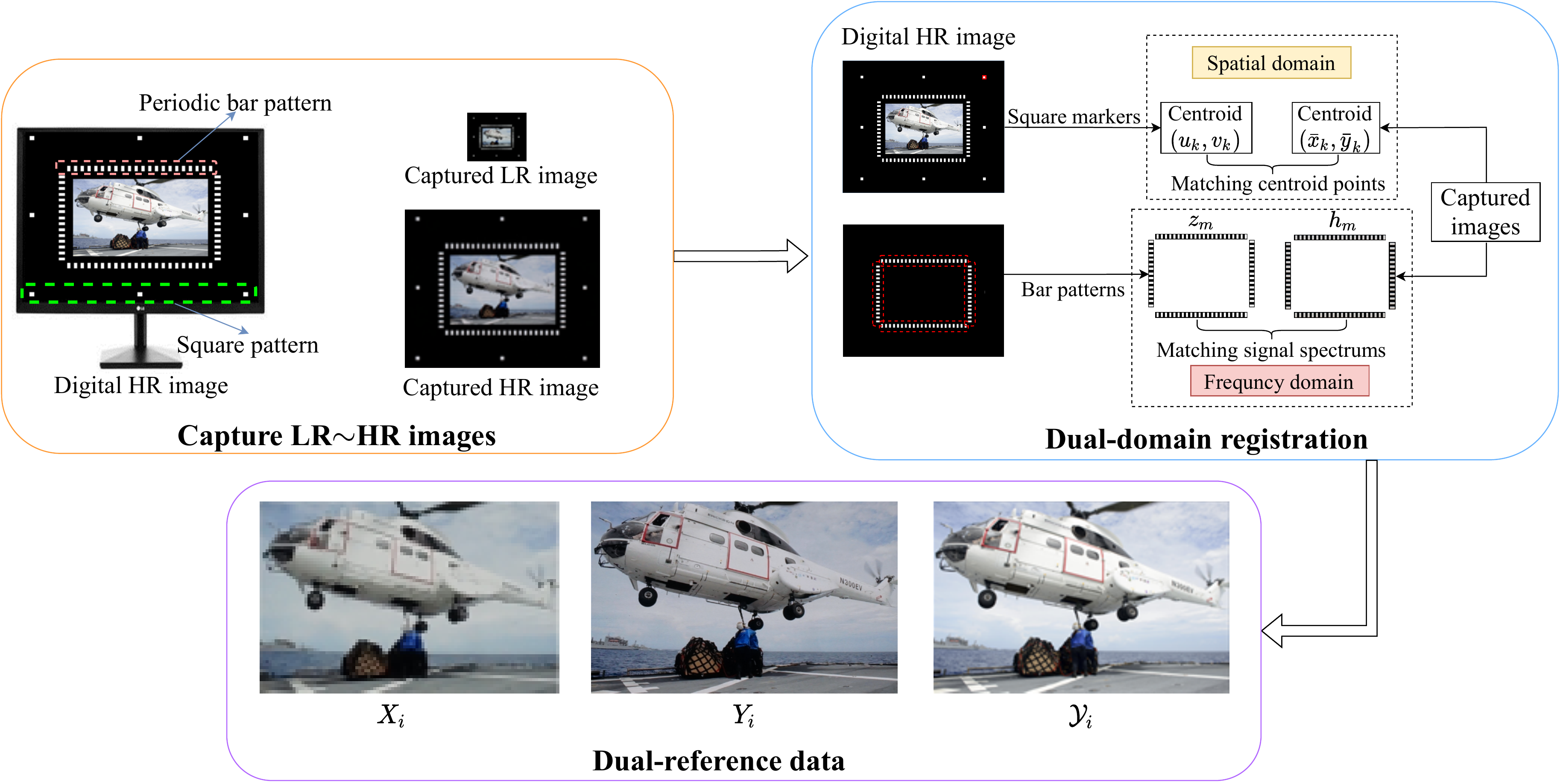}
   \caption{Overall process of the dual-domain image registration.
   The captured LR image $X_i$ and the captured HR image $Y_i$ are aligned to the digital image $\mathcal{Y}_i$.}
   \label{Figure7_rigistration}
\end{figure*}

\section{DRTI Image Acquisition System}\label{DRTI_system}

In this section, we present the details of the proposed process of generating SR training datasets of
monitor-induced dual-reference paired images (DRTI).

\begin{table}[h!]
   \centering
   \caption{Cameras and their settings for collecting DRTI image triplets.}
   \scalebox{0.95}{

    \begin{tabular}{||>{\centering\arraybackslash}m{.2\linewidth}|>{\centering\arraybackslash}m{.2\linewidth}|>{\centering\arraybackslash}m{.2\linewidth}|>{\centering\arraybackslash}m{.2\linewidth}||}
    \hline
   Camera & Sony Nex-6 & Canon 5D-Mark\uppercase\expandafter{\romannumeral3} & Sony Alpha-a7R\uppercase\expandafter{\romannumeral2} \\ [0.5ex]
   \hline\hline
   Aperture & f/10 & f/10 & f/10 \\
   Shutter speed & 1/30s & 1/50s & 1/30s \\
   Focal length & 18mm\&80mm & 30mm\&120mm & 18mm\&80mm \\
   Pixel pitch  & 4.8${\rm\mu m}$ & 6.2${\rm\mu m}$ &  6.0${\rm\mu m}$\\
   Pixel area & 22.75${\rm \mu m}^2$ &38.69${\rm \mu m}^2$ &20.25${\rm \mu m}^2$\\
   Distance & 5.3m& 6.2m& 4.3m\\[0.5ex]
   \hline\hline
   Image triplets & 600 & 600 & 600 \\
   \hline
   \end{tabular}

   }
   \label{table_camera}
\end{table}

\subsection{System hardware and setup}
We collect DRTI data by shooting natural images displayed on a 4K UHD (3840$\times$2160) HDR10 monitor by digital cameras, and obtain LR images with short focal lens and HR images with long focal lens.
To prevent blur caused by possible vibrations of the camera, we mount the camera on a track dolly slider and trigger the shutter using a WiFi remote controller.
Three cameras, Sony Nex-6, Canon 5D-Mark\uppercase\expandafter{\romannumeral3} and Sony Alpha-a7R\uppercase\expandafter{\romannumeral2}, are used to increase the variety of camera statistics in reality.

The screen shooting is carried out in a controlled dark room lab.\ environment to eliminate ambient lights. The camera sensor projection plane is calibrated to parallel with and have sufficient
distance from the screen surface to prevent moir\'e interferences and geometric distortions.
Denote the focal length of the camera, the object distance and the pixel pitch of the screen by $f$, $d$ and $u_s$ respectively.  The size of an imaged
pixel on screen through the lens is

\begin{equation}
   u_i = u_s\frac{f}{d}
   \label{lens_eq}
\end{equation}

\noindent Moir\'e patterns are caused by the interference between the screen image and the color filter array of the camera.
By the Nyquist–Shannon sampling theorem, moir\'e patterns can be eliminated if
the camera senor sampling step size $u_c>2u_i$.  Then it follows from Eq.~\ref{lens_eq} that the moir\'e-free distance between the camera and the screen is

\begin{equation}
   d > 2f\frac{u_s}{u_c}
   \label{camera_pos}
\end{equation}

\noindent As the focal length for capturing LR images is shorter than the focal length for HR images, eliminating moir\'e patterns in captured HR images will also ensure no moir\'e patterns in LR images.  Therefore we determine the camera position by the long focal length for capturing HR images.
In addition, to simplify the LR$\sim$HR registration, we set the focal plane (sensor array) of the camera parallel to the screen surface.
To achieve this we display a square pattern on the screen, and make sure the pattern to remain a square in the captured screen image, by adjusting the universal angle plate on which the camera is mounted.

\subsection{Data collection process}
We conduct multiple shooting sessions using three cameras to ensure a sufficiently large data volume
and diversity to facilitate deep learning. The camera specifications and settings for the data acquisition process are tabulated in
Table~\ref{table_camera}. To keep the consistency of each captured LR$\sim$HR image pair $(X_i,Y_i)$,
we only change the focal length of the lens once for each camera. More specifically, we first capture all needed LR images with
the same short focal length, then we adjust the focal length and capture the corresponding HR images.
This can reduce the registration error caused by the frequent mechanical changes.

To remove all unnecessary variations among acquired images, all camera parameters irrelevant to spatial resolution, such as white balance, aperture size and exposure time, should be fixed.
We thus collect raw Bayer images rather than JPEG images produced by the camera built-in image processing pipeline. The LR and HR raw Bayer images are demosaiced into corresponding RGB images of linear scale without any tone mapping nor gamma correction.
We correct the lens distortion with Zhang's method \cite{zhang2000flexible}.
To factor out the influences of monitor backlight, we display a black image on the monitor at the beginning of every shooting process and subtract it from every captured image afterwards.
The ultra-high quality monitor and the DSLR camera are remotely operated in a dark room so as to prevent any interference of ambient light and any motions of the monitor and camera.

\section{LR$\sim$HR Registration}\label{sec_image_resig}
In the DRTI acquisition system, the original images $\mathcal{Y}_i$ are displayed on a flat monitor to be captured into LR and HR pairs.  This flatness eliminates the possibility of LR$\sim$HR misalignment due to depth disparity, which is inevitable when capturing real scenes \cite{zhang2019zoom, cai2019toward}.
In addition, after the pre-calibration of making the camera image plane parallel to the monitor surface, the registration of the LR$\sim$HR image pair is simplified to an affine transformation.

\subsection{Centroid based spatial registration}

To determine the affine transformation, at least three paired keypoints are required.
The paired keypoints are typically determined by finding and matching unique feature points in the scene by the well-known methods such as SIFT \cite{lowe2004distinctive} and SURF \cite{bay2006surf}.
But distinctive sharp features are not always available.
To ensure robust and highly accurate LR$\sim$HR registration,
we design and display registration markers on screen margins as shown in the left of Fig.~\ref{Figure7_rigistration}. These markers consist of a solid white square against black background.  In this design, the centroid of the white square pattern is of subpixel precision and it is resistant to sensor noises in image acquisition.  Moreover, the centroid is
invariant to scaling, rotation and translation of the image.  Thanks to these properties we can use the centroid of the white square marker to pursue high registration precision.

We use eight solid white square markers, four at the corners of the image, and the other four at the center of the four sides of the image.
These eight markers are for LR$\sim$HR registration in spatial domain.
To register the dual-reference training images, we align the captured LR image $X_i$ and the captured HR image $Y_i$ to the displayed digital image $\mathcal{Y}_i$, respectively.
To keep the training images in the same resolution, we upsample $X_i$ and downsample $\mathcal{Y}_i$ to the size of $Y_i$.  The registration process for $(X_i,\mathcal{Y}_i)$ is the same for $(Y_i,\mathcal{Y}_i)$.

Fig~\ref{Figure7_rigistration} illustrates the overall registration process for DRTI.
For each white square marker, we calculate its centroid as an anchor point for image alignment.
Denoting the image segment of the $k$-th square marker by $A_k(x,y)$,
the $k$-th anchor point of the LR image is
\begin{equation}
(\bar{x}_k,\bar{y}_k) = (\frac{\sum\limits_{A_k} x A_k(x,y)}{|A_k|},
\frac{\sum\limits_{A_k} y A_k(x,y)}{|A_k|})
\end{equation}
where $|A_k|$ is sum of all pixel values in marker segment $A_k$.

Denote the alignment affine transformation matrix by $\left[\begin{matrix}\mathbb{S},{\mathbf{b}}\end{matrix}\right]$,
where $\mathbb{S} \in \mathbb{R}^{2\times2}$ is for rotation and scaling, and $\mathbf{b}\in \mathbb{R}^{2\times1}$ is the translation vector.
The eight anchor points $(\bar{x}_k,\bar{y}_k)$, $1 \leq k \leq 8$, of the LR image
are aligned to the corresponding anchor points $(u_k,v_k)$ of the digital HR image $\mathcal{Y}_i$
by minimizing the following objective function $f_1(\mathbb{S},{\mathbf{b}})$ in homogeneous coordinate representation:

\begin{equation}
   \underset{\mathbb{S},{\mathbf{b}}}{\min}
   \left\| \begin{bmatrix} \mathbb{S} & {\mathbf{b}}\\
   \mathbf{0} & 1  \end{bmatrix}
   \underbrace{
   \begin{bmatrix}
   \bar{x}_1 & \cdots  & \bar{x}_8   \\
   \bar{y}_1 & \cdots  & \bar{y}_8  \\
   1 & \cdots  & 1    \\
   \end{bmatrix}
   }_{A}
   -
   \underbrace{
   \begin{bmatrix}
   u_1 & \cdots  & u_8   \\
   v_1 & \cdots  & v_8  \\
   1 & \cdots  & 1    \\
   \end{bmatrix}
   }_{B}
   \right\|_F
   \label{obj_func1}
\end{equation}
whose solution is   $\left[\begin{matrix}\mathbb{S}^*,\mathbf{b}^*\end{matrix}\right] = BA^{{\scriptscriptstyle \prime}}(AA^{{\scriptscriptstyle \prime}})^{-1}$.

\subsection{Dual-domain registration}

\begin{figure*}[t]
   \centering
  \begin{subfigure}[t]{0.32\textwidth}
     \centering
     \includegraphics[width=1\textwidth]{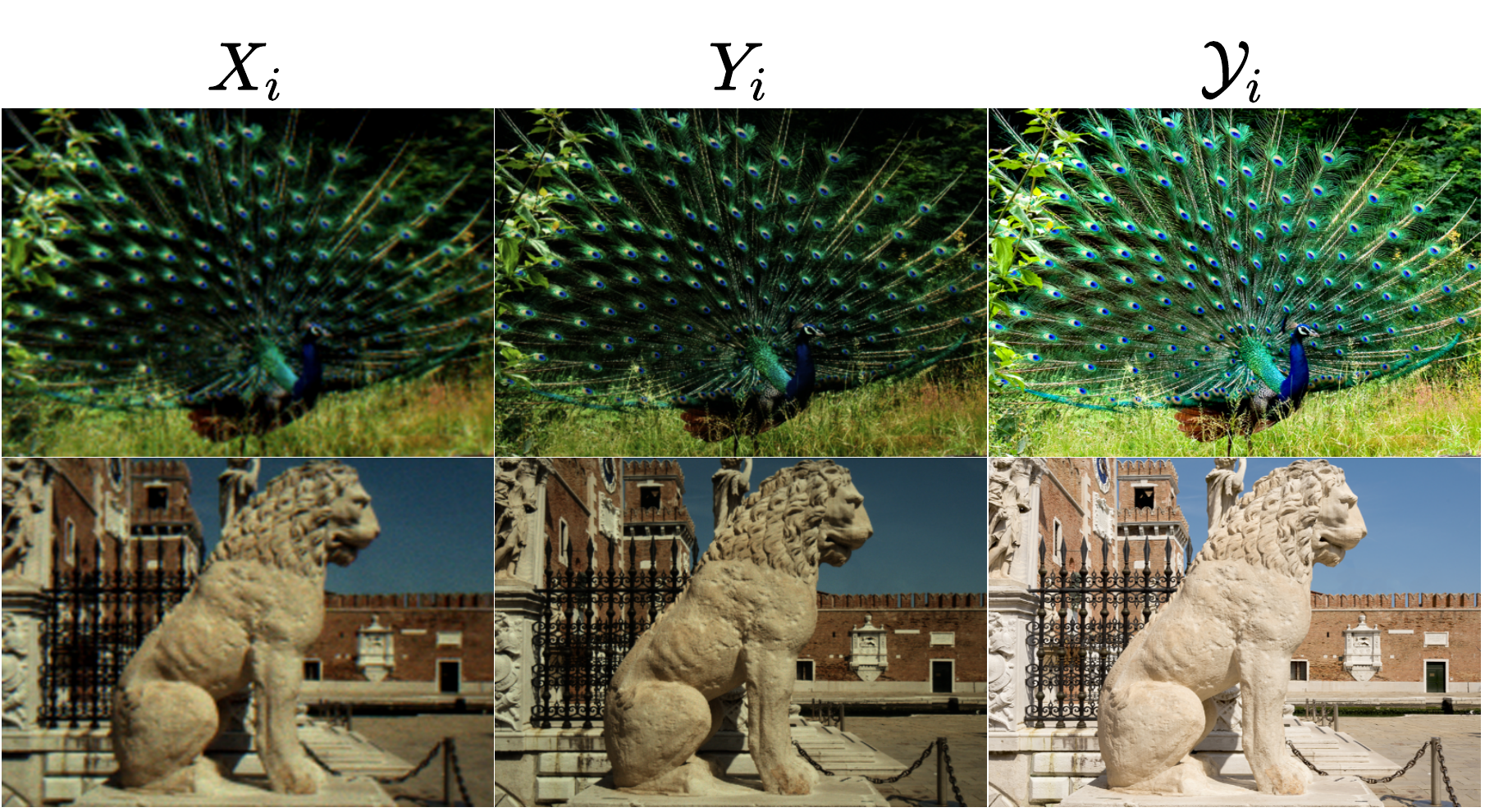}
     \caption*{Sony Nex}
  \end{subfigure}
  \begin{subfigure}[t]{0.32\textwidth}
     \centering
     \includegraphics[width=1\textwidth]{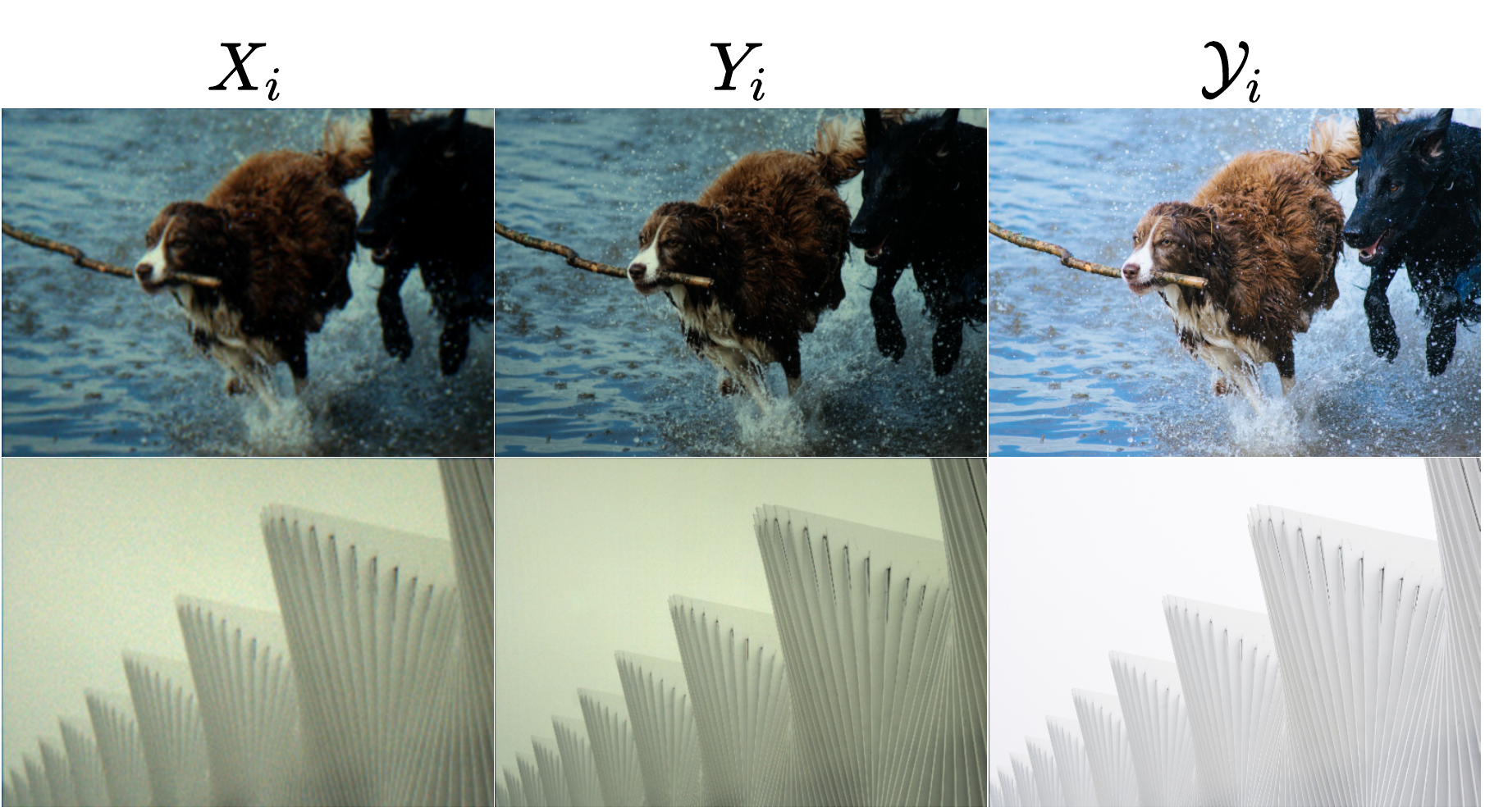}
     \caption*{Canon 5D}
  \end{subfigure}
  \begin{subfigure}[t]{0.32\textwidth}
     \centering
     \includegraphics[width=1\textwidth]{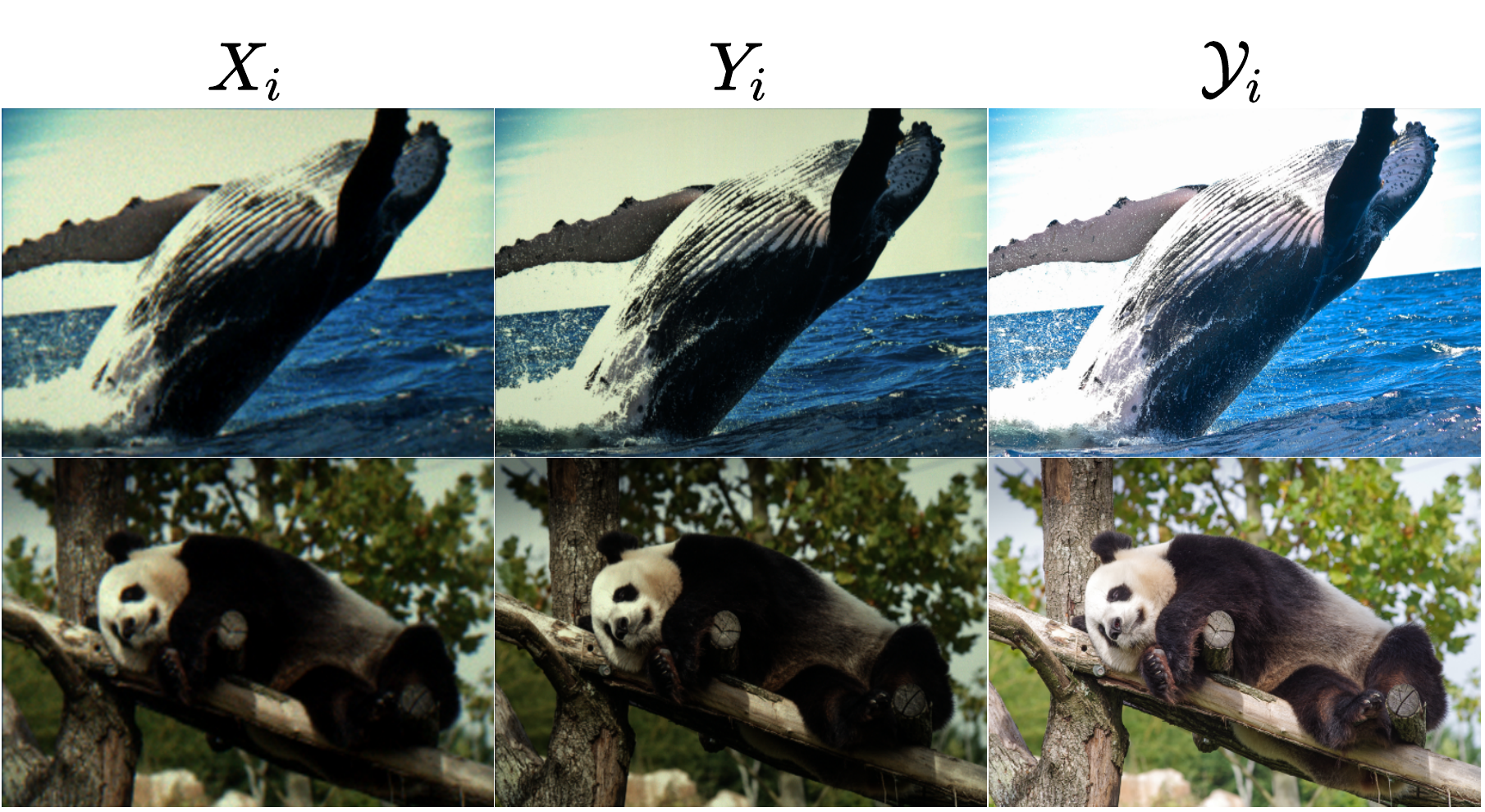}
     \caption*{Sony Alpha}
  \end{subfigure}

  \caption{Samples of DRTI image triplets $(X_i, Y_i, \mathcal{Y}_i)$ collected by three different cameras.}
  \label{fig:sample}
\end{figure*}

The precision of the above centroid based spatial LR$\sim$HR registration can be improved by the following process of joint alignment in both frequency and spatial domains.
Specifically, we design a periodic bar pattern of a fixed frequency and display it along the four sides of the screen, as shown in the right of Fig~\ref{Figure7_rigistration}.
With its peak frequency and orientation known, the bar pattern offers a very strong prior, \ie, a singular point in the frequency domain.
This known prior allows us to apply the affine theorem in frequency domain \cite{Bracewell1993AffineTF} to achieve very high registration precision even amidst noises in image acquisition.

However, as we will soon see, true alignment in frequency domain requires solving a non-convex optimization problem.
The global optimal solution can be guaranteed only if the algorithm is launched from an initial point that is sufficiently close to the global optimum.
The required initialization is the role of the centroid based affine transform. This is the rationale for our dual-domain approach to LR$\sim$HR registration.

For the dual-domain registration, we extract the four paired regular bar patterns from the initial spatially aligned image pair.
Then we refine the affine transformation matrix $\left[\begin{matrix}\mathbb{S},{\mathbf{b}}\end{matrix}\right]$ jointly in both frequency and spatial domains.  Let the captured bar pattern segment be $h_m$, $1 \leq m \leq 4$. Given a pixel position $\mathbf{x} =(x_m,y_m)$ in the bar pattern segment, we have
\begin{equation}
   z_m(\mathbf{x}) = h_m(\mathbb{S}\mathbf{x} + \mathbf{b})
   \label{affine_1}
\end{equation}

According to the affine theorem of the frequency domain \cite{Bracewell1993AffineTF}, we can rewrite Eq.~\ref{affine_1} as

\begin{equation}
   \mathcal{Z}_m(\mathbf{w}) = \frac{1}{\left |\mathrm{det}(\mathbb{S})\right |}e^{j2\pi \mathbf{b}^{\prime}\mathbb{S}^{\dagger}\mathbf{w}}\mathcal{H}_m(\mathbb{S}^{\dagger}\mathbf{w})
\end{equation}
where $\mathcal{H}_m(\mathbf{w})$ and $h_m(\mathbf{x})$ are the Fourier-transform pair of the captured bar pattern segment; $\mathcal{Z}_m(\mathbf{w})$ and $z_m(\mathbf{x})$ are the Fourier-transform pair of the digital bar pattern segment; the symbol $\dagger$ stands for the inverse of a transposed matrix, $\mathbb{S}^{\dagger} = (\mathbb{S}')^{-1}$.
As there is no flipping transformation between the image pair, we can have $\mathrm{det}(\mathbb{S}) >0$.

By design, the ground truth phase $\theta_m$ and the ground truth frequency $\mathbf{w}_m$ of the periodic registration bar pattern are known.
The difference $\Delta_m$ between $z_m$ and the ground truth signal in frequency domain is
\begin{equation}
   \begin{aligned}
   \Delta_m = \frac{\mathcal{Z}_m(\mathbf{w}_m)}{e^{ j2\pi \theta_m}} =
   \frac{1}{\mathrm{det}(\mathbb{S})}e^{j2\pi (\mathbf{b}^{\prime}\mathbb{S}^{\dagger}\mathbf{w}_m-\theta_m)}\mathcal{H}_m(\mathbb{S}^{\dagger}\mathbf{w}_m)
   \end{aligned}
   \label{affine_2}
\end{equation}

In the frequency domain, if the phase difference of two identical periodic signals is made zero,
then the two signals are perfectly aligned in phase in frequency domain or having no translation in spatial domain.
Besides, if the two signals achieve the maximum magnitude of the frequency response at the designed frequency point $\mathbf{w}_m$, then
the two signals are perfectly aligned in frequency domain, or having no scaling nor rotation errors in spatial domain.
To achieve both the phase difference minimization and the magnitude maximization, it amounts to maximizing the real part of $\Delta_m$ for each bar pattern segment.  For robust global alignment, the phase registration process takes into account all four periodic registration bar patterns, two horizontal and two vertical, and calls for solving the following optimization problem:
\begin{equation}
   \begin{aligned}
   \underset{\mathbb{S},\mathbf{b}}{\max} f_2(\mathbb{S},\mathbf{b})= \underset{\mathbb{S},\mathbf{b}}{\max} \sum\limits_{m=1}^{4}\operatorname{Re}(\Delta_m)
   \end{aligned}
   \label{affine_3}
\end{equation}
where $\operatorname{Re}(\cdot)$ represents the real part of the complex number.

To ensure the optimal solution in both spatial and frequency domains, we carry out the following dual-domain joint optimization:
\begin{equation}
   \underset{\mathbb{S},\mathbf{b}}{\min}\,f_1(\mathbb{S},\mathbf{b}) - \beta f_2(\mathbb{S},\mathbf{b})
   \label{affine_joint}
\end{equation}
where $\beta$ is the Lagrangian multiplier.
Starting from the initialization by the centroid based spatial registration,
a gradient descent algorithm can find the optimal alignment transform
$[\mathbb{S}^*,\mathbf{b}^*]$.
Using $[\mathbb{S}^*,\mathbf{b}^*]$, we align the captured LR image $X_i$ with the digital image $\mathcal{Y}_i$.  The same dual-domain registration algorithm is used to align the captured HR image $Y_i$ with the digital image $\mathcal{Y}_i$.
The proposed dual-domain registration method can achieve super-high precision, producing the shift error, on average, below 0.25 pixel, the scale and rotation errors below $1\times 10^{-4}$ and  $1\times 10^{-3}$, respectively.
Fig~\ref{fig:sample} presents six registered dual-reference samples of DRTI.

 \begin{figure}[h]
   \centering
   \includegraphics[width=0.48\textwidth]{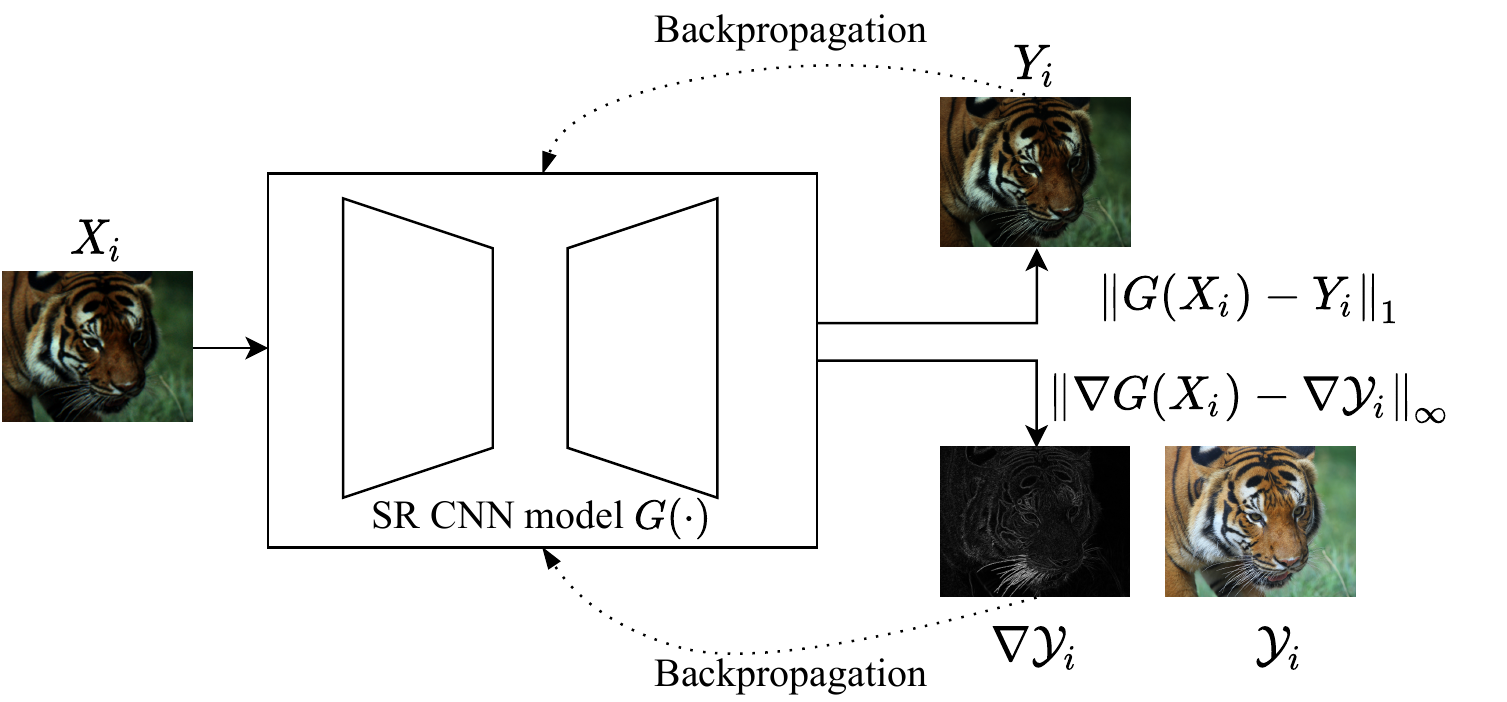}
   \caption{Dual-reference supervised learning with DRTI image triplets $(X_i,Y_i,\mathcal{Y}_i)$.}
   \label{Figure_training}
 \end{figure}
\section{Dual-Reference Learning}\label{sec_loss_func}

After being aligned, the DRTI image triplets $(X_i,Y_i,\mathcal{Y}_i)$ are ready for use in the supervised learning of the DCNN SR models.  The LR$\sim$HR pair $(X_i,Y_i)$ provides the information on camera physical attributes, such as pixel PSF, sensor noise statistics, lens characteristics, \etc., while
the original HR digital image $\mathcal{Y}_i$ provides structural
information on high frequency features that might be compromised by the analog operation of screen shooting.

Whatever the DCNN architecture we choose for the SR task, we supervise the SR model training by both the captured $Y_i$ and the digital file $\mathcal{Y}_i$.
This calls for a dual-reference loss function:
\begin{equation}
   \resizebox{.9\hsize}{!}{$\mathcal{L}(G(X_i),Y_i,\mathcal{Y}_i)=\left\|G(X_i) -Y_i\right\|_1 + \lambda \left\|\nabla G(X_i) - \nabla \mathcal{Y}_i\right\|_{\infty}$}
   \label{dual_loss_func}
\end{equation}
where $G(\cdot)$ represents an arbitrary SR CNN model, $\nabla$ is the differential operator and $\lambda$ is the trade-off weight.
The former $\ell_1$-norm term minimizes the pixel-wise errors between the prediction $G(X_i)$ and the captured HR image $Y_i$.

\begin{figure*}[htbp]
   \centering
   \includegraphics[width=1\textwidth]{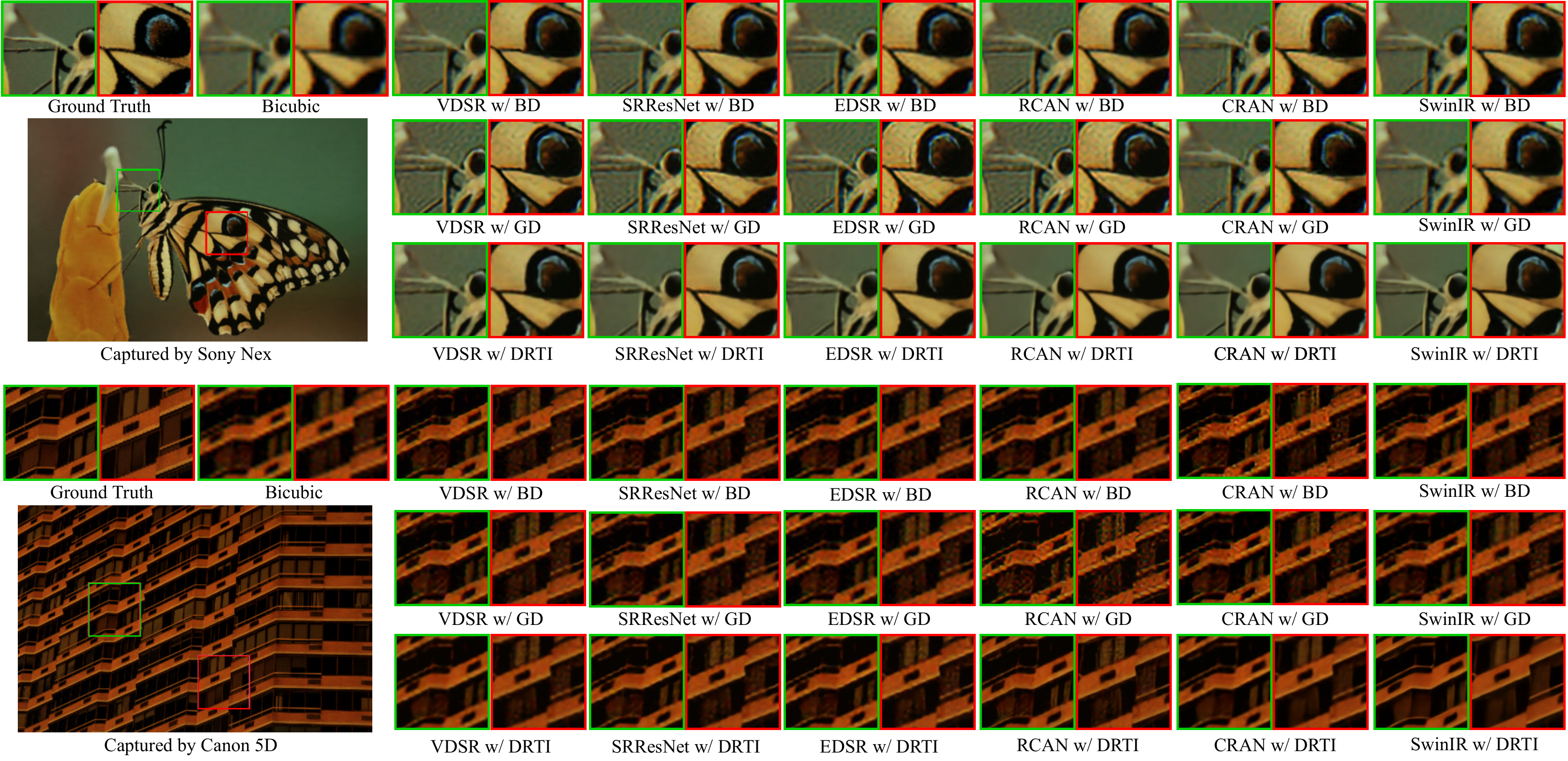}
   \caption{Results of different DCNN SR models trained on synthetic datasets and the DRTI dataset.}
   \label{Figure8_Screen_Comparison}
 \end{figure*}

In addition, we introduce a regulation term to enhance the gradient similarity \cite{Mehdi2021,upsample_sa2008,yu2012robust} between the prediction $G(X_i)$ and the digital HR image $\mathcal{Y}_i$,
improving the high frequency textures of the prediction $G(X_i)$.
To avoid producing over-smooth results, we adopt an infinite norm.
By minimizing the dual-reference loss function $\mathcal{L}(G(X_i),Y_i,\mathcal{Y}_i)$,
the pixel PSF between $X_i$ and $Y_i$ and the high frequency information of $\mathcal{Y}_i$ can be jointly
learned by the SR CNN model in an end-to-end manner, as illustrated in Fig~\ref{Figure_training}.

\begin{table*}[ht]
   \centering
   \caption{PSNR and SSIM values of different DCNN SR models trained by synthetic and camera-captured DRTI datasets.}
   \scalebox{1}{
   \begin{tabular}{||c|c|c|c|c|c|c|c|c||}
   \hline\hline
   \multirow{2}{*}{\bfseries Methods (x4)} & \multicolumn{2}{c|}{\bfseries Original} & \multicolumn{2}{c|}{\bfseries BD} & \multicolumn{2}{c|}{\bfseries GD} & \multicolumn{2}{c||}{\bfseries DRTI}\\
   \cline{2-9}
          & PSNR & SSIM &PSNR & SSIM & PSNR &SSIM & PSNR &SSIM \\
   \hline\hline
   Bicubic  &  29.26  & 0.811	 &29.26  & 0.811	  &   29.26   &	0.811   & 29.26  & 0.811\\
   \hline\hline
   VDSR\cite{kim2016accurate}    & 29.67& 0.813 &30.41 & 0.838    &   30.64   &	0.841  &	 \textbf{32.12}  &  \textbf{0.860}  \\\hline
   SRResNet\cite{ledig2017photo}  &	29.89& 0.824 &30.51   & 0.837	   & 30.59     & 0.836  & \textbf{32.53}   & \textbf{0.871}   \\\hline
   EDSR\cite{lim2017enhanced}   & 29.86& 0.824 &30.68   & 	0.838    &   30.65   &	0.838  &	\textbf{32.57}  &  \textbf{0.871}  \\\hline
   RCAN\cite{zhang2018image}   & 29.99	& 0.827  & 30.78  &   0.842    &   30.71   &	0.839   &	 \textbf{33.15}  &  \textbf{0.877}  \\\hline
   CRAN\cite{zhang2021context}  & 29.96& 0.823 & 30.85    &  0.844   & 30.77   & 0.840  &	 \textbf{33.13}  &  \textbf{0.878}  \\\hline
   SwinIR\cite{liang2021swinir} & 30.01& 0.832 & 30.81  &   0.849    &  30.75  &	0.844 &	 \textbf{33.18}  &  \textbf{0.883}  \\\hline
   \hline
   \end{tabular}
   }
   \label{table_method_comparison}
   % \vspace{-0.65cm}
\end{table*}

\begin{figure*}[htbp]
   \centering
   \setlength{\abovecaptionskip}{0cm}
   \setlength{\belowdisplayskip}{-0.1cm}
   \includegraphics[width=0.86\textwidth]{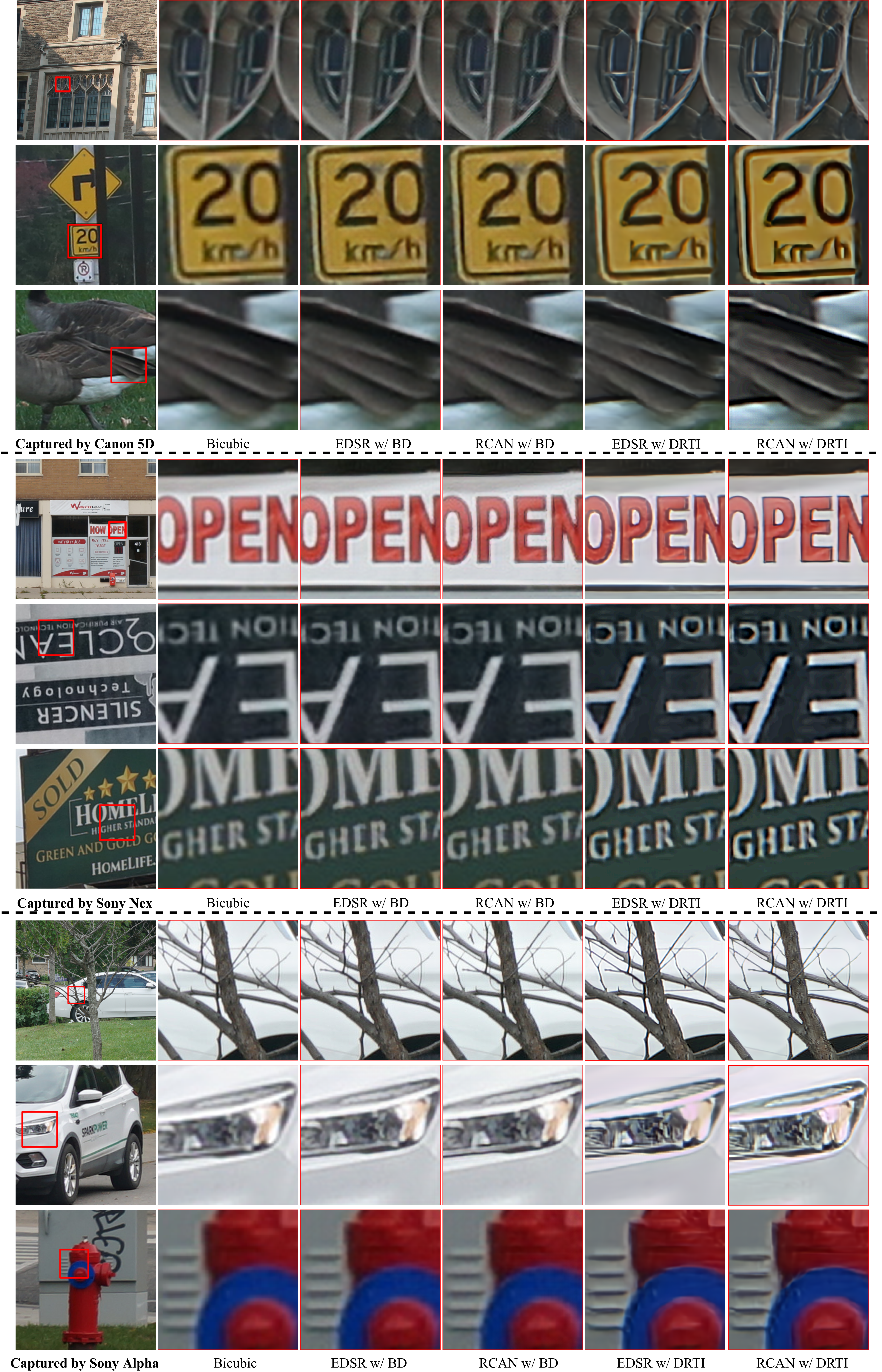}
   \caption{Results of $\times 4$ superresolving real-world images captured by three DSRL cameras.  Note improvements in visual quality brought by training EDSR and RCAN with the DRTI datasets over training with synthetic BD datasets.}
   \label{Figure9_real_comparison}
   % \vspace{-0.5cm}
\end{figure*}

\begin{figure*}[htbp]
   \centering
   \begin{subfigure}[t]{0.31\textwidth}
      \includegraphics[width=1.02\textwidth]{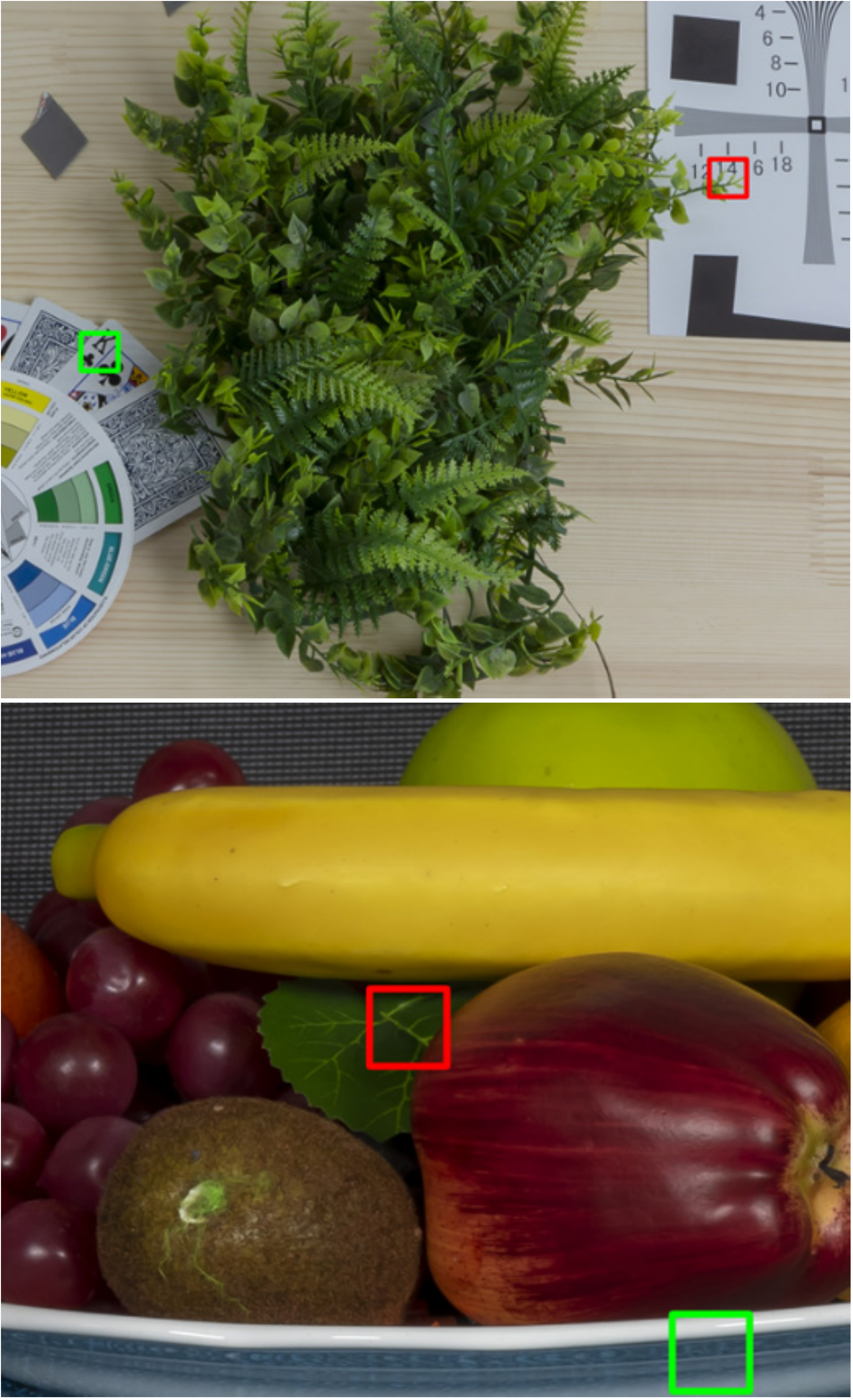}
      \caption*{\scriptsize Test images \cite{cai2019toward}}
   \end{subfigure}%\\
   \hspace{0.12cm}
   % \begin{minipage}[b]{0.68\textwidth}
   \begin{subfigure}[t]{0.13\textwidth}
      \includegraphics[width=1\textwidth]{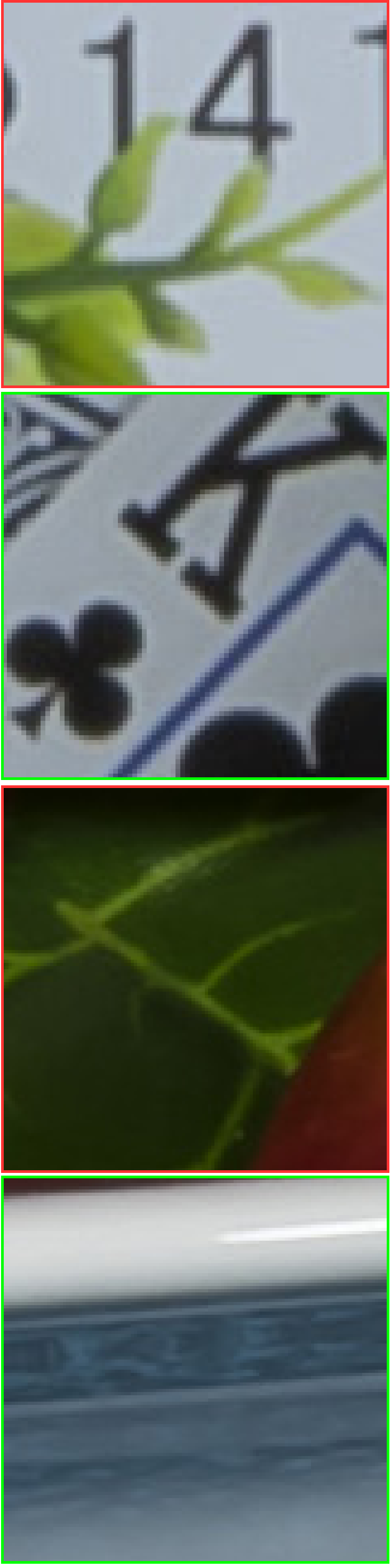}
      \caption*{\scriptsize Ground Truth}
   \end{subfigure}
   \hfill
   \begin{subfigure}[t]{0.13\textwidth}
      \includegraphics[width=1\textwidth]{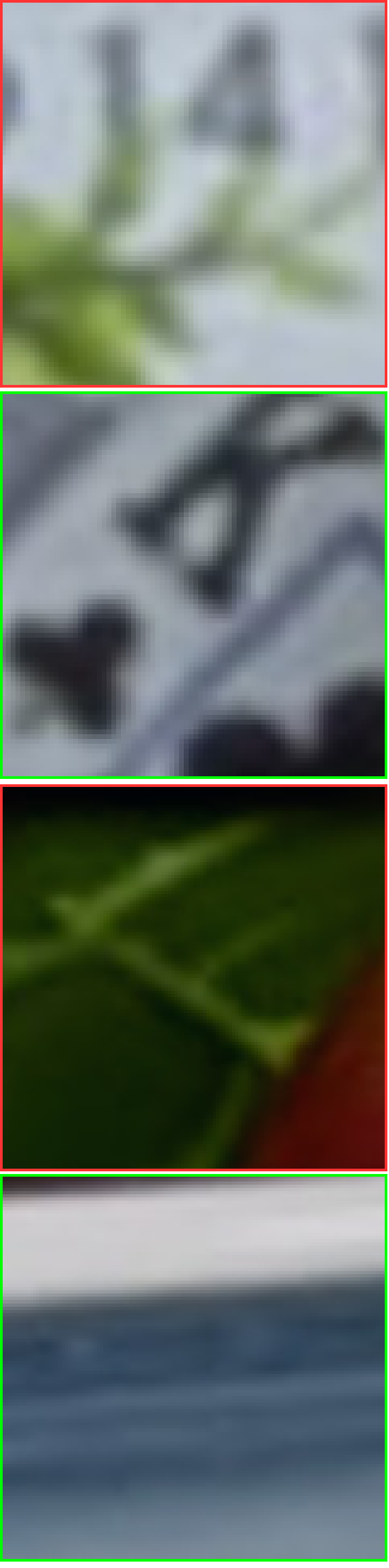}
      \caption*{\scriptsize Bicubic}
   \end{subfigure}
   \hfill
   \begin{subfigure}[t]{0.13\textwidth}
      \includegraphics[width=1\textwidth]{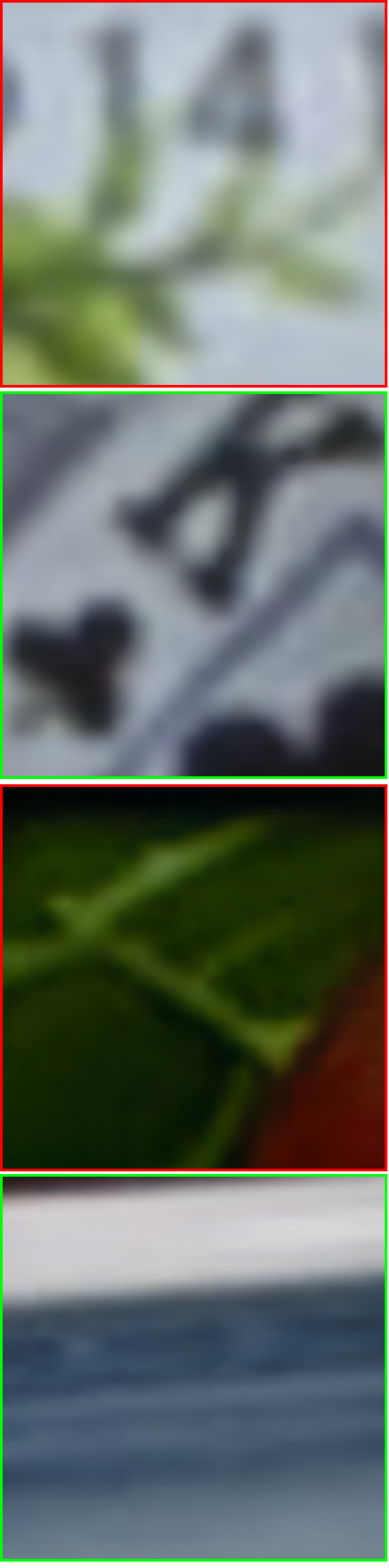}
      \caption*{{\scriptsize RCAN w/ Chen's\cite{chen2019camera}}}
   \end{subfigure}
   \hfill
   \begin{subfigure}[t]{0.13\textwidth}
      \includegraphics[width=1.0\textwidth]{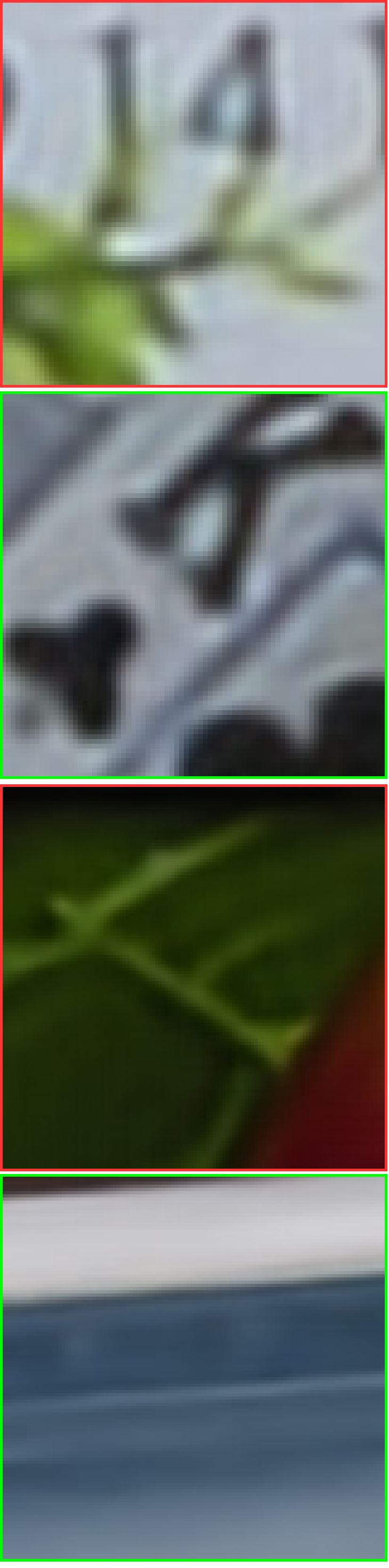}
      \caption*{{\scriptsize RCAN w/ Cai's\cite{cai2019toward}}}
   \end{subfigure}
   \begin{subfigure}[t]{0.13\textwidth}
      \includegraphics[width=1\textwidth]{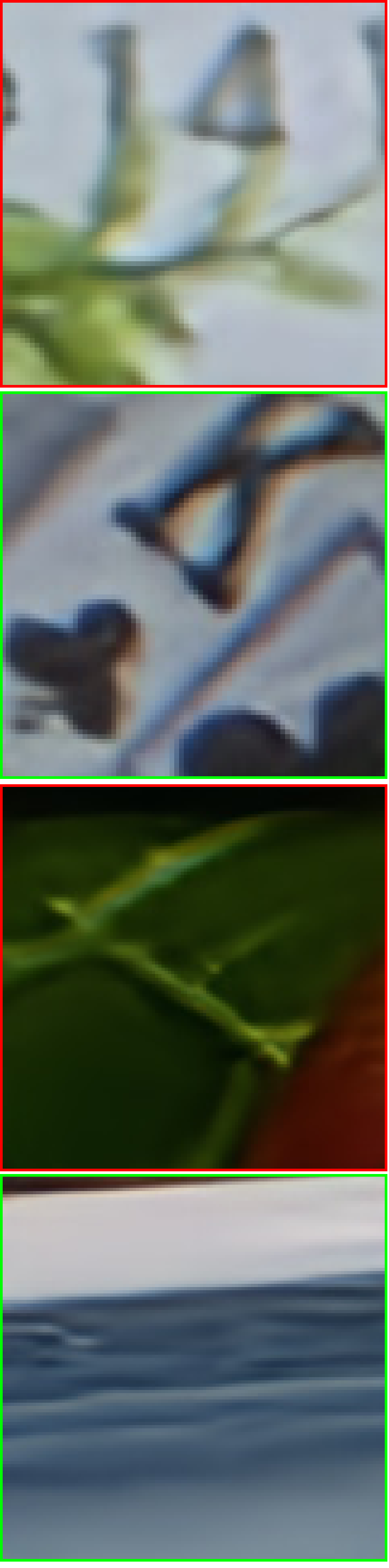}
      \caption*{{\scriptsize RCAN w/ DRTI}}
   \end{subfigure}%\\
   \caption{SR results of training RCAN by different camera-captured datasets but inferring on the LR images of Cai's\cite{cai2019toward} dataset.}
   \label{Figure10_test_cai}
   % \vspace{-0.5cm}
\end{figure*}

\section{Experimental Results}\label{results}

\subsection{Design and settings of experiments}

As listed in Table.~\ref{table_camera}, we use three cameras (Sony Nex-6, Canon 5D-Mark, Sony Alpha-a7R) to collect three DRTI datasets denoted by $S_1,S_2,S_3$, each containing 600 image triples $(X_i,Y_i,\mathcal{Y}_i)$, where the 600 displayed images are
selected from the training set of DIV2K \cite{timofte2017ntire}.  For comparison purposes, we also generate two synthetic datasets of LR$\sim$HR pairs, denoted by $S_4$ and $S_5$; the LR images in $S_4$ (or $S_5$) are
synthesized by bicubic (or Guassian) downsampling of the corresponding captured HR images.  We use the paired dataset $S_j$ to train a CNN super-resolution model $M_j$, $1 \leq j \leq 5$.
Nine-tenth of image pairs in $S_j$ are used for training $M_j$ and validation, and the remaining one-tenth for inference tests.

In the training process, we randomly crop the training images into $256 \times 256$ patches to train all the models and the mini-batch size in all the experiments is set to 8. The trade-off weight $\lambda$ is set 0.1.
Adam optimizer \cite{kingma2014adam} is chosen to train all models by setting initial learning rate $10^{-4}$. The learning rate is fixed in first
10K iterations and linearly decay to $10^{-6}$ after 70K iterations. With same parameters setting, all the experiments are conducted on a PC with single NVIDIA TITAN XP GPU.  Following the mainstream practice in the field \cite{lim2017enhanced,zhang2018image}, quantitative metrics PSNR and SSIM are calculated on the Y channel in the YCbCr space for quantitative performance evaluations.

CNN methods are very sensitive to nuance differences between the training and inference data. Therefore, a CNN SR method will suffer performance losses if it is trained by images taken by one camera but is used to restore images taken by another.  Needless to say, model $M_j$ performs the best on images in $S_j$.  But we are more interested in the far harder task of using $M_j$ to infer on images in $S_k$, $k \not = j$; namely, we evaluate the generalization capabilities of the models trained by DRTI datasets in comparison with the models trained by the synthetic datasets and other captured datasets.

Because degradations produced by the camera lens and sensors vary from camera to camera, the DRTI-trained models are not device-free. To assess the impacts of the camera dependency, we also conduct cross-camera tests.

\subsection{Results on laboratory datasets}
The ultimate criterion of success for the proposed DRTI data collection procedure is whether the DRTI-trained DCNN model for super resolution can indeed outperform the counterpart trained by other LR$\sim$HR datasets when both being applied to infer on real-world images.
But for the sake of completeness and reference purpose, let us first evaluate the inference results on DRTI data.  We compare how a given DCNN SR model performs when it is trained by different datasets, including both synthesized and DRTI.
In the comparison study, two synthetic LR$\sim$HR datasets, denoted by BD and GD, are generated by downsampling the captured HR images $Y_i$ of the DRTI dataset using bicubic and Gaussian kernels.
To fairly evaluate the difference between the synthetic image paris and the real image pairs of the captured DRTI dataset, in this experiment, all DRTI-trained DCNN models are supervised by
the captured image pairs $(X_i,Y_i)$ without $\mathcal{Y}_i$.
Table.~\ref{table_method_comparison} reports the PSNR and SSIM values (with scaling factor $\times 4$) of different SR models VDSR\cite{kim2016accurate}, SRResNet\cite{ledig2017photo}, EDSR\cite{lim2017enhanced}, RCAN\cite{zhang2018image}, CRAN\cite{zhang2021context} and SwinIR\cite{liang2021swinir}
after trained by different datasets.
As a reference, we also report the performance of these models trained by the original DIV2K dataset (denoted by ``Original" in Table.~\ref{table_method_comparison} ).

As reported in Table.~\ref{table_method_comparison}, for each of the seven tested SR methods, the DRTI-trained model has significant improvement over the BD-trained model and the GD-trained model, with gains from 1.7dB to 2.4dB.
The visual quality comparison is presented in Fig.~\ref{Figure8_Screen_Comparison}. As shown, the SR results associated with the training datasets BD and GD have more artifacts and more blurred edges than with the DRTI dataset.
It is expected that our DRTI dataset outperforms BD and GD, because it characterizes statistical relations between the LR and HR images more accurately for the camera.

\begin{figure*}[!htp]
   \centering
   % \vspace{-0.3cm}
   % \setlength{\abovecaptionskip}{-0.01cm}
   \begin{subfigure}[t]{0.195\textwidth}
      \includegraphics[width=1.0\textwidth]{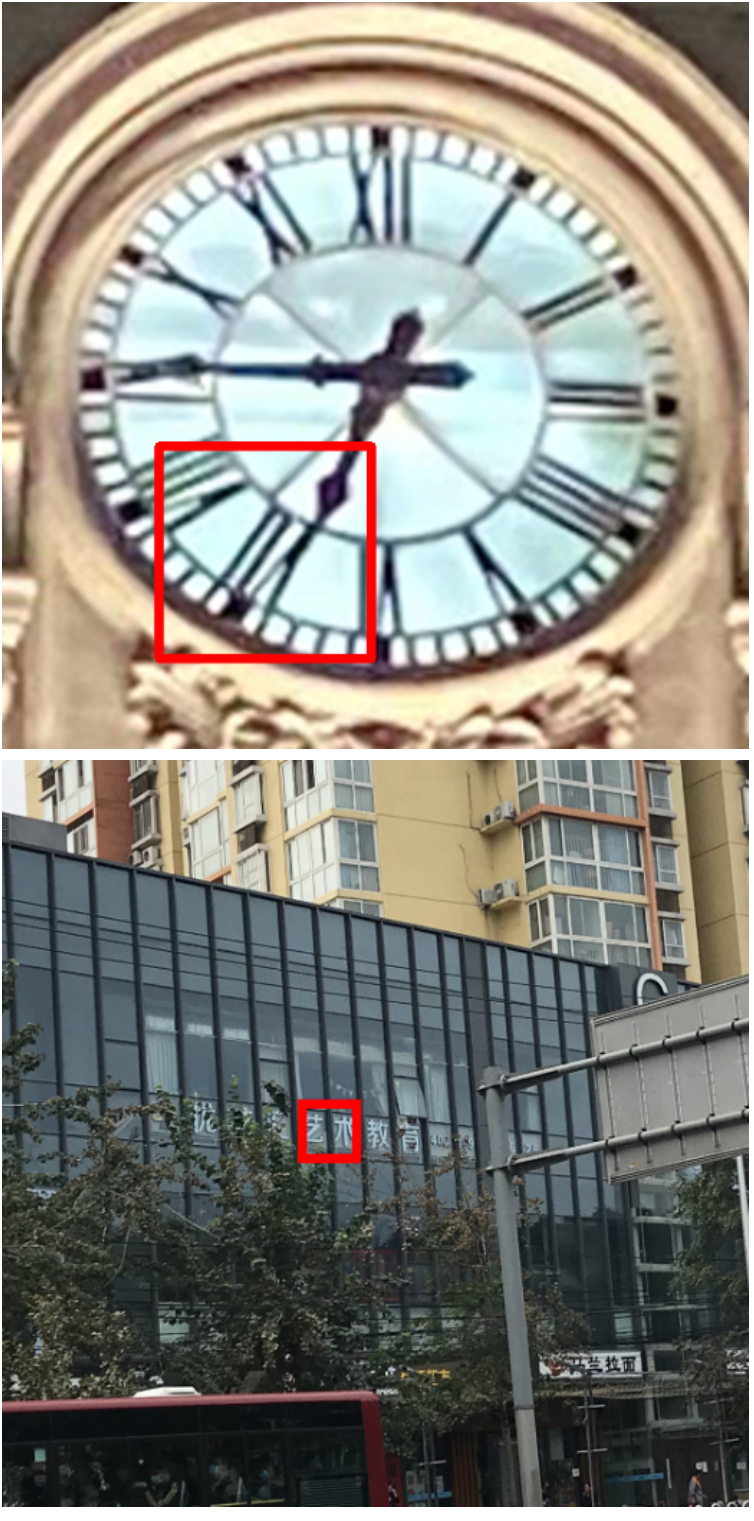}
      \caption*{{Captured by iPhone 7plus}}
   \end{subfigure}%\\
   \hspace{0.01cm}
   % \begin{minipage}[b]{0.68\textwidth}
   % \hfill
   \begin{subfigure}[t]{0.195\textwidth}
      \includegraphics[width=1.0\textwidth]{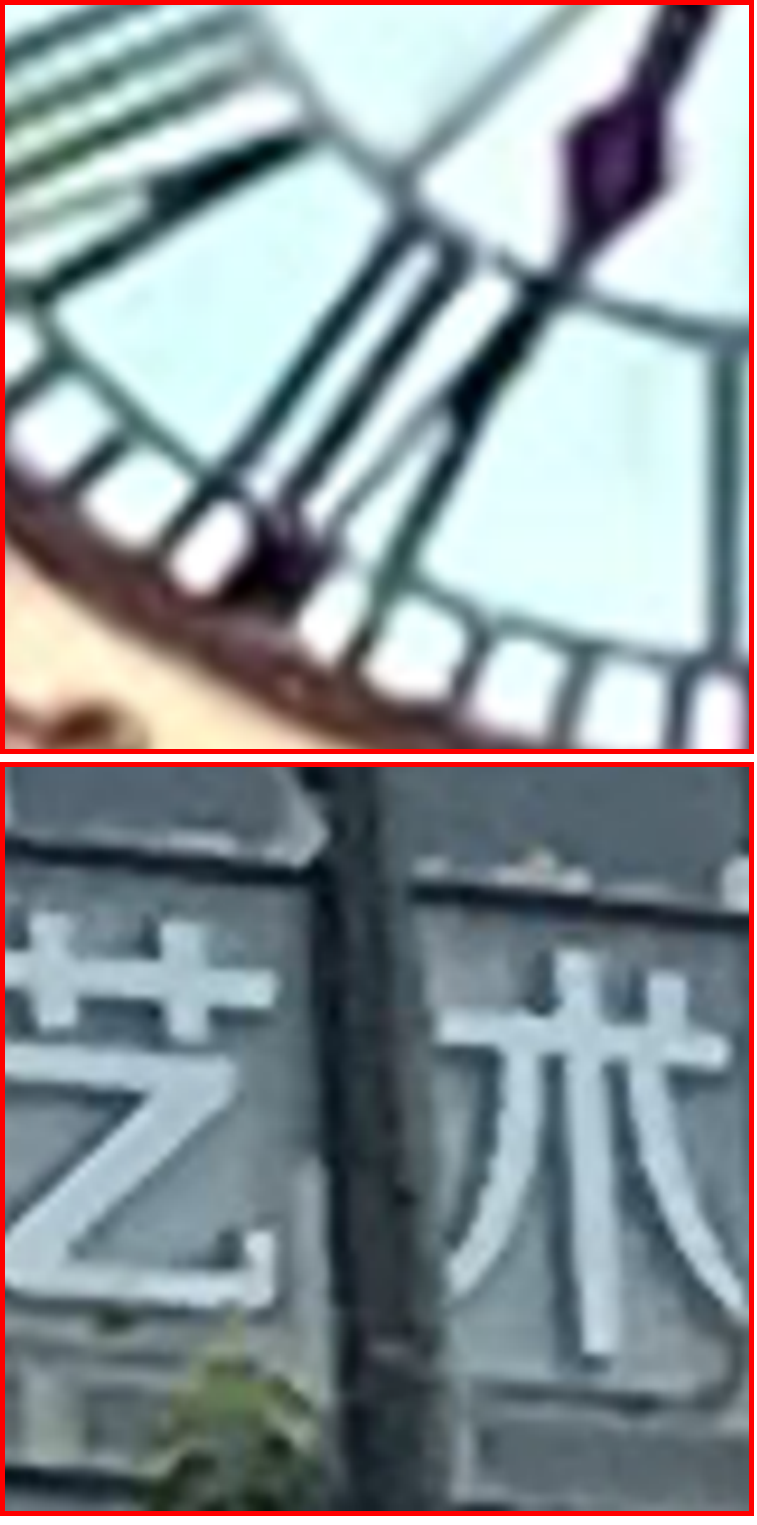}
      \caption*{Bicubic}
   \end{subfigure}
   \hfill
   \begin{subfigure}[t]{0.195\textwidth}
      \includegraphics[width=1.0\textwidth]{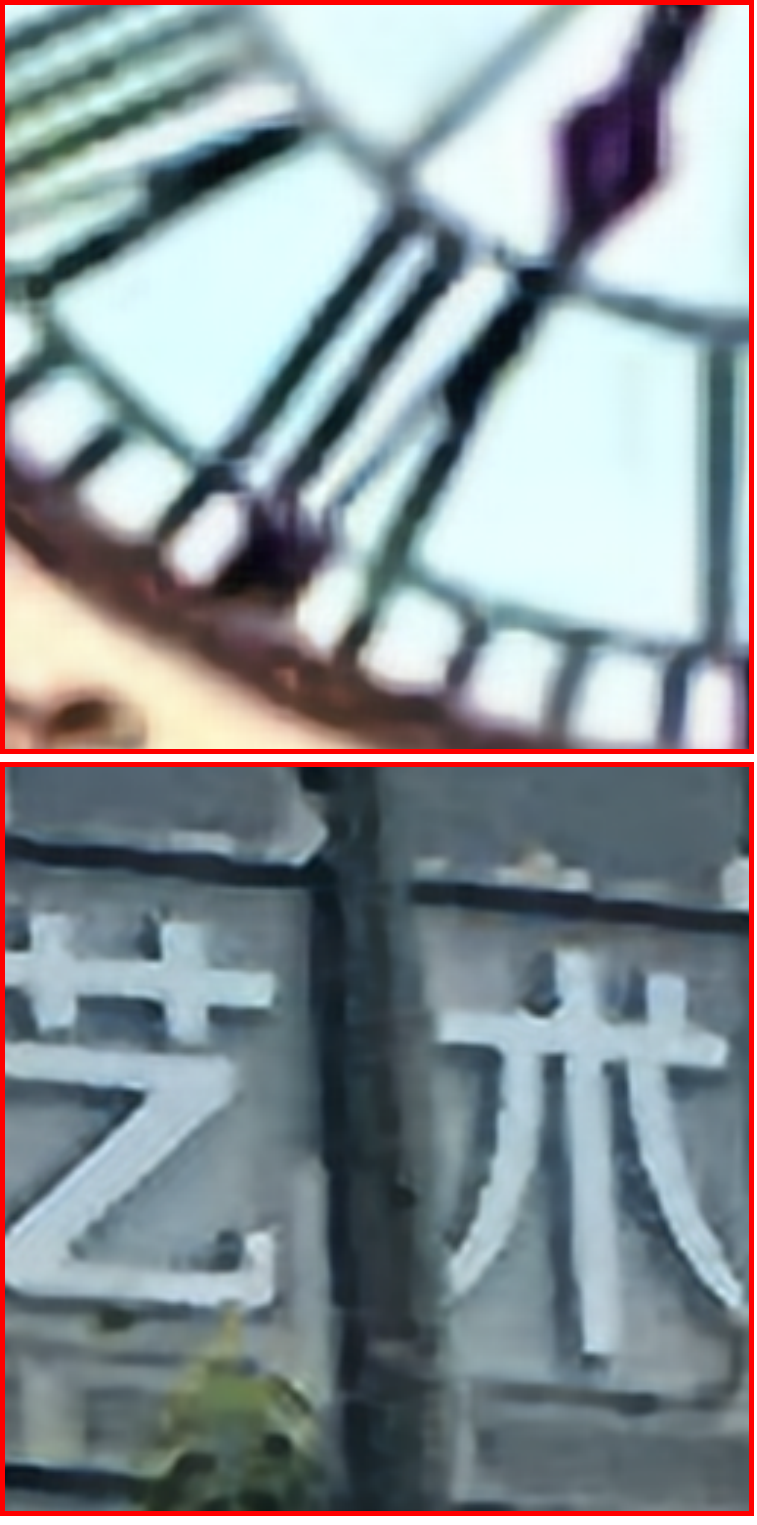}
      \caption*{{ RCAN w/ Chen's\cite{chen2019camera}}}
   \end{subfigure}
   \hfill
   \begin{subfigure}[t]{0.195\textwidth}
      \includegraphics[width=1.0\textwidth]{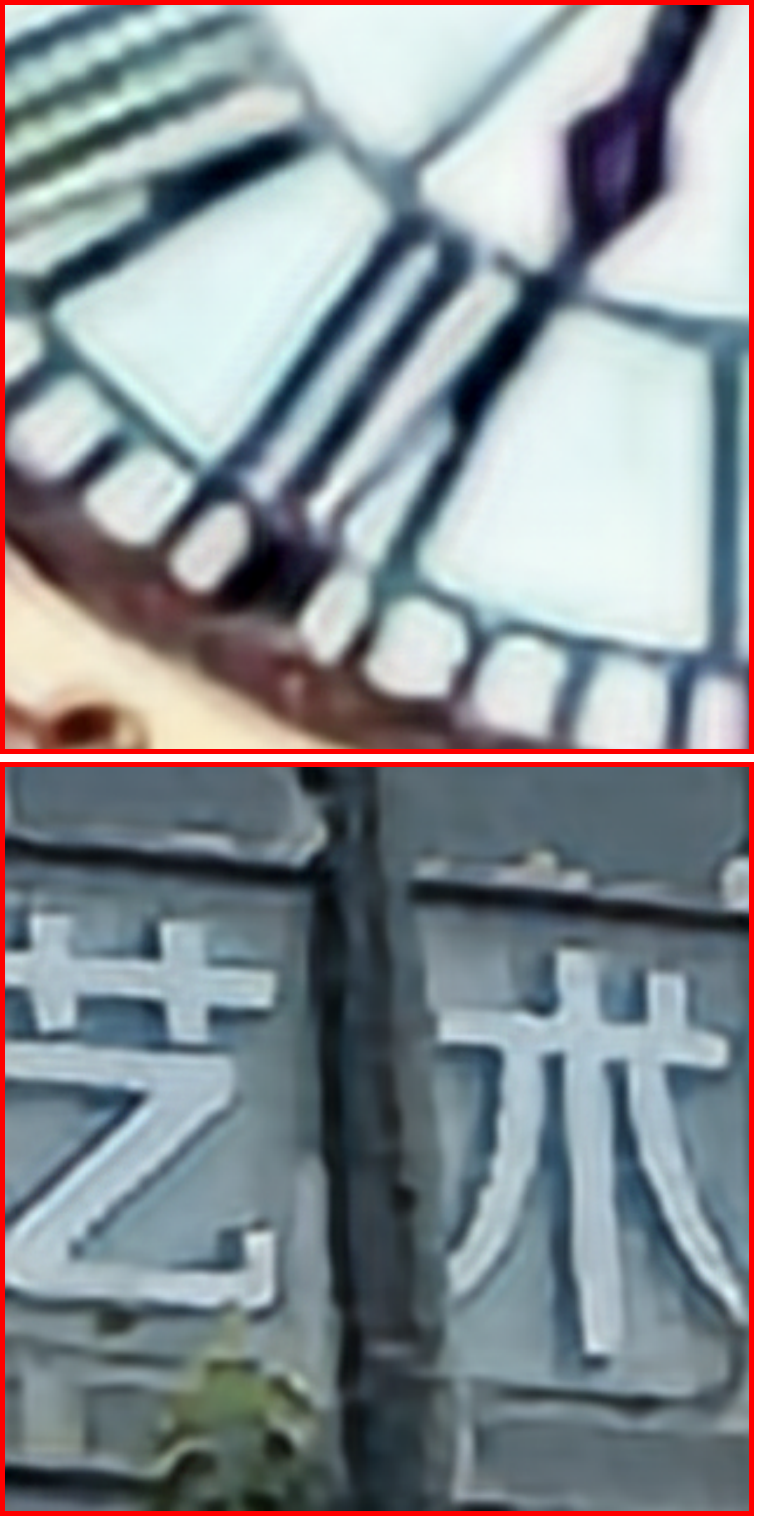}
      \caption*{{ RCAN w/ Cai's\cite{cai2019toward}}}
   \end{subfigure}
   \hfill
   \begin{subfigure}[t]{0.195\textwidth}
      \includegraphics[width=1.0\textwidth]{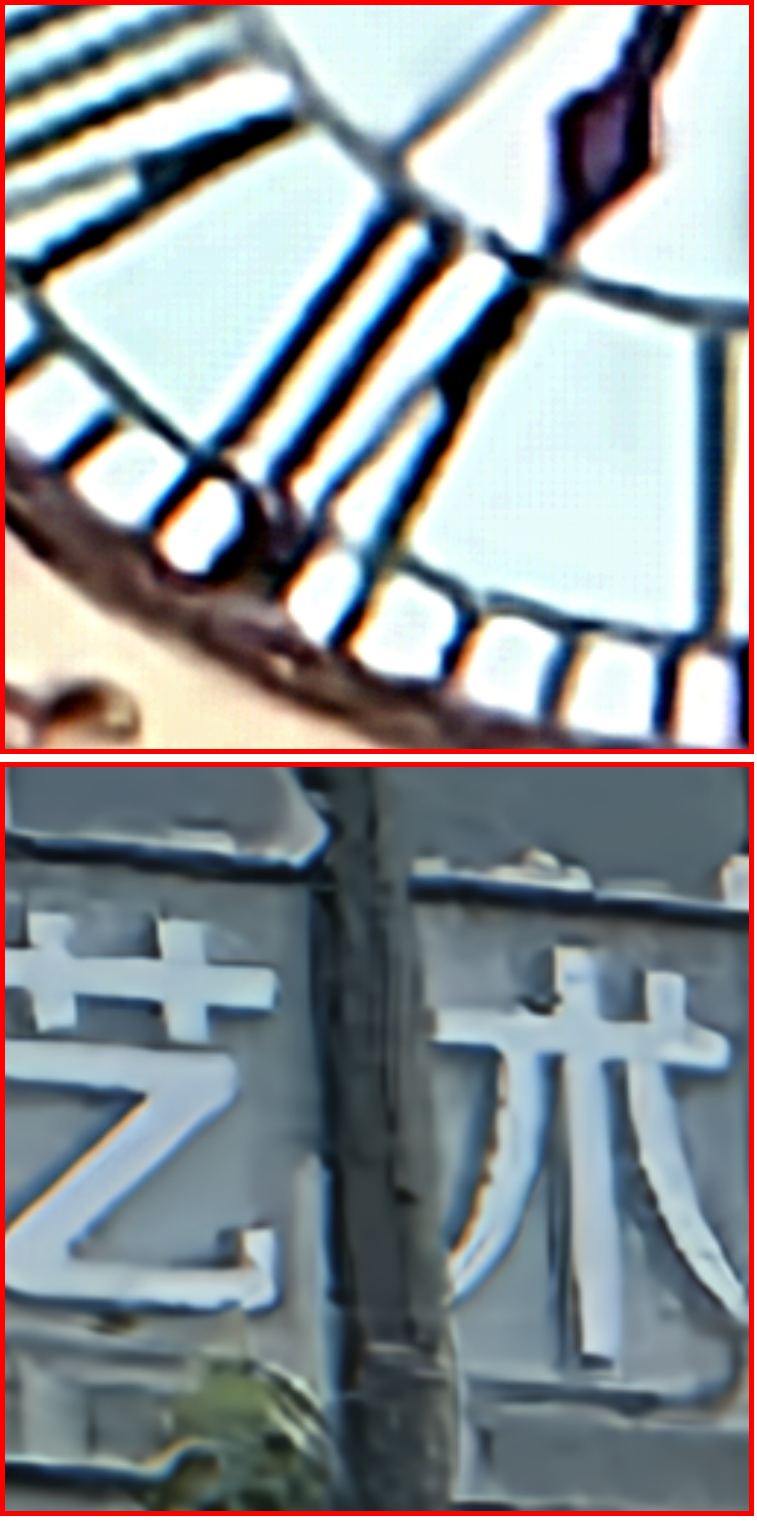}
      \caption*{{ RCAN w/ DRTI}}
   \end{subfigure}\\
   \vspace{0.4cm}
   \begin{subfigure}[t]{0.195\textwidth}
      \includegraphics[width=1.0\textwidth]{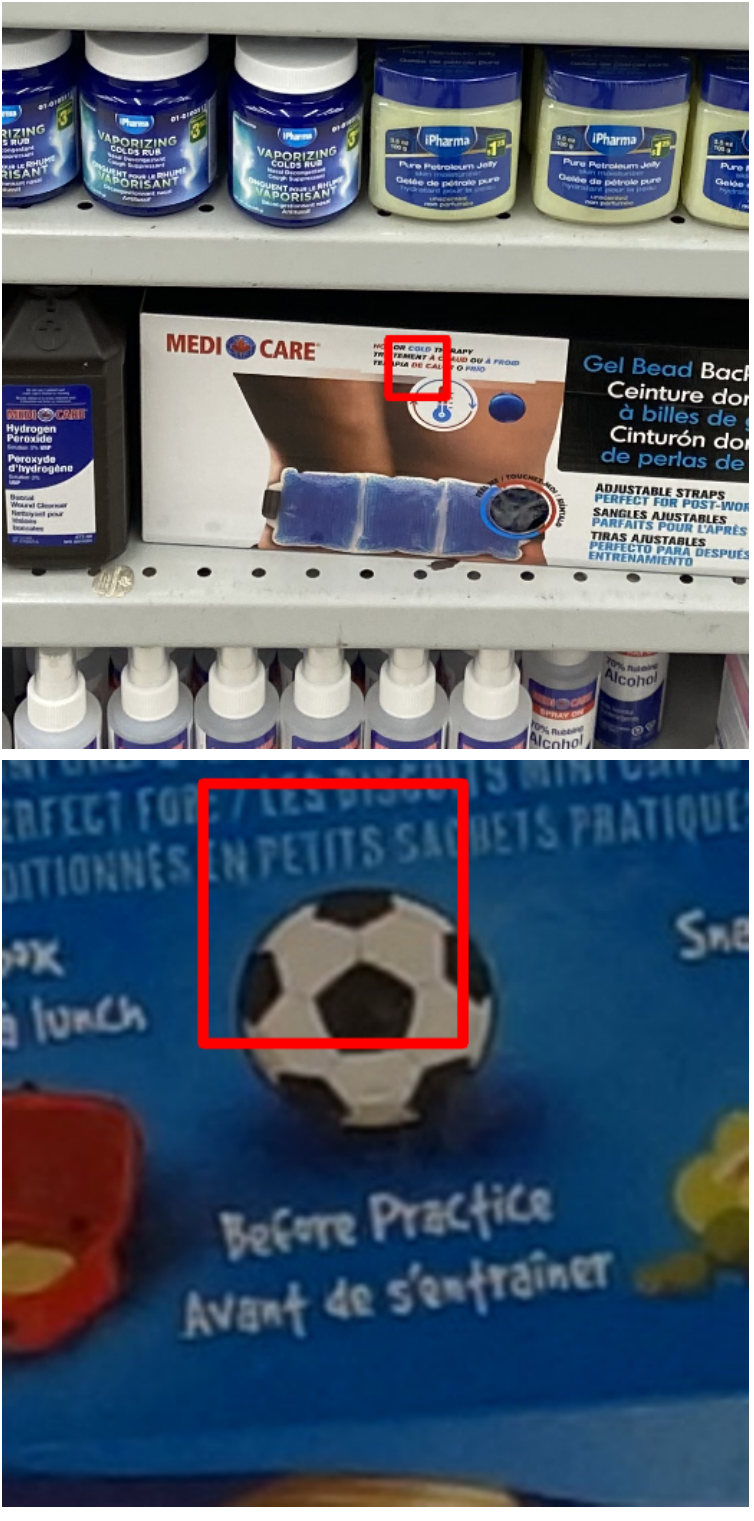}
      \caption*{{ Captured by iPhone 11}}
   \end{subfigure}%\\
   \hspace{0.01cm}
   % \begin{minipage}[b]{0.68\textwidth}
   % \hfill
   \begin{subfigure}[t]{0.195\textwidth}
      \includegraphics[width=1.0\textwidth]{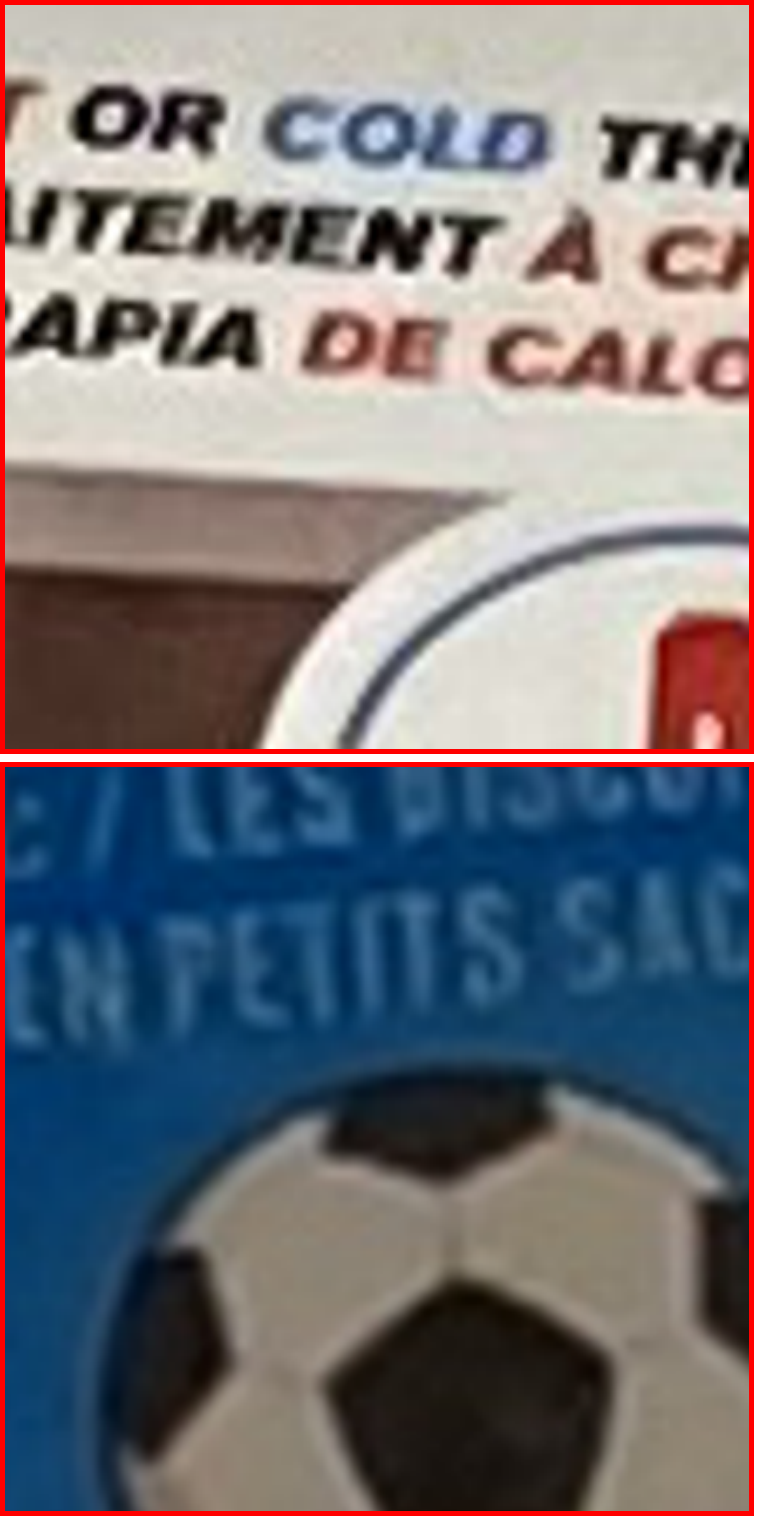}
      \caption*{Bicubic}
   \end{subfigure}
   \hfill
   \begin{subfigure}[t]{0.195\textwidth}
      \includegraphics[width=1.0\textwidth]{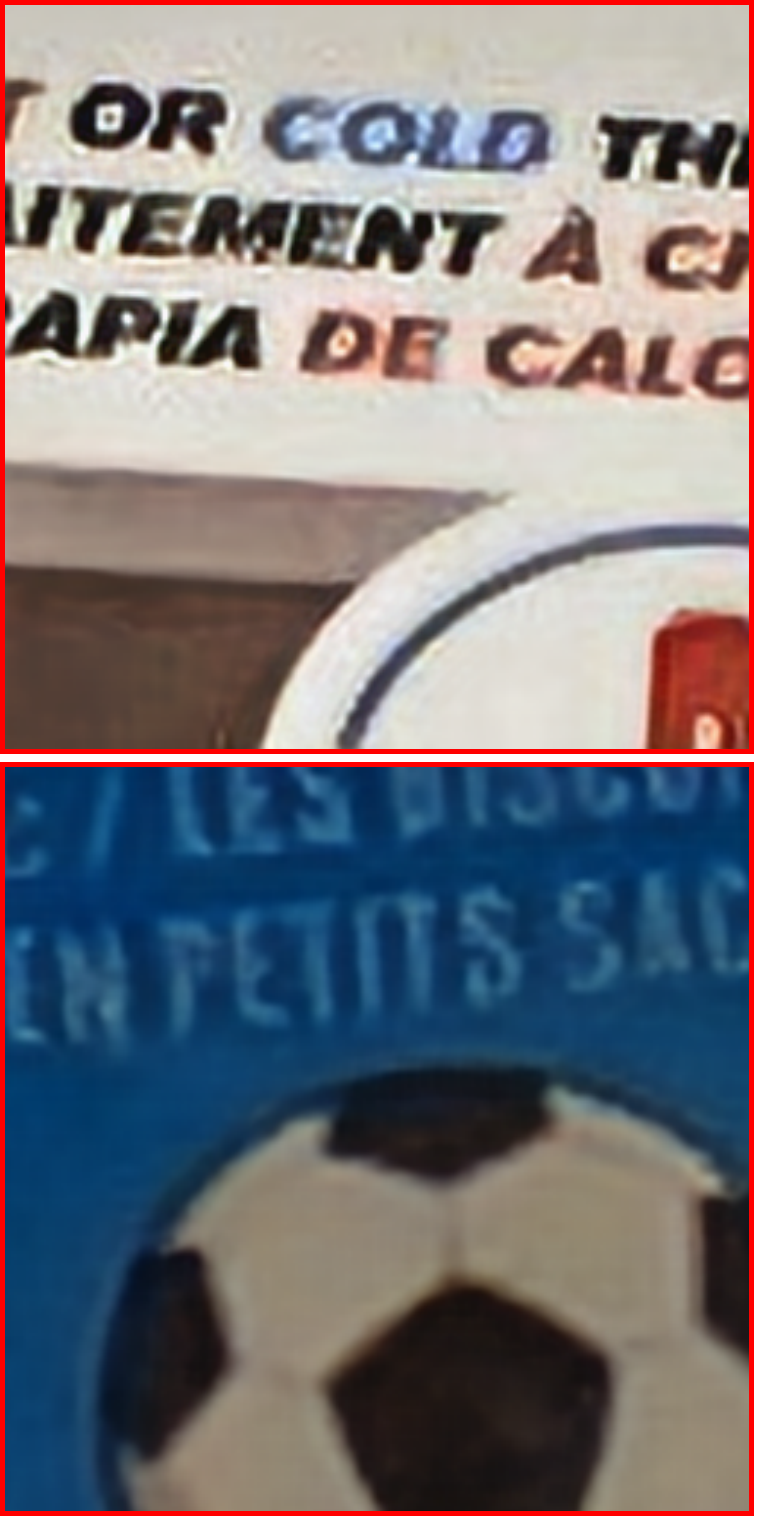}
      \caption*{{RCAN w/ Chen's\cite{chen2019camera}}}
   \end{subfigure}
   \hfill
   \begin{subfigure}[t]{0.195\textwidth}
      \includegraphics[width=1.0\textwidth]{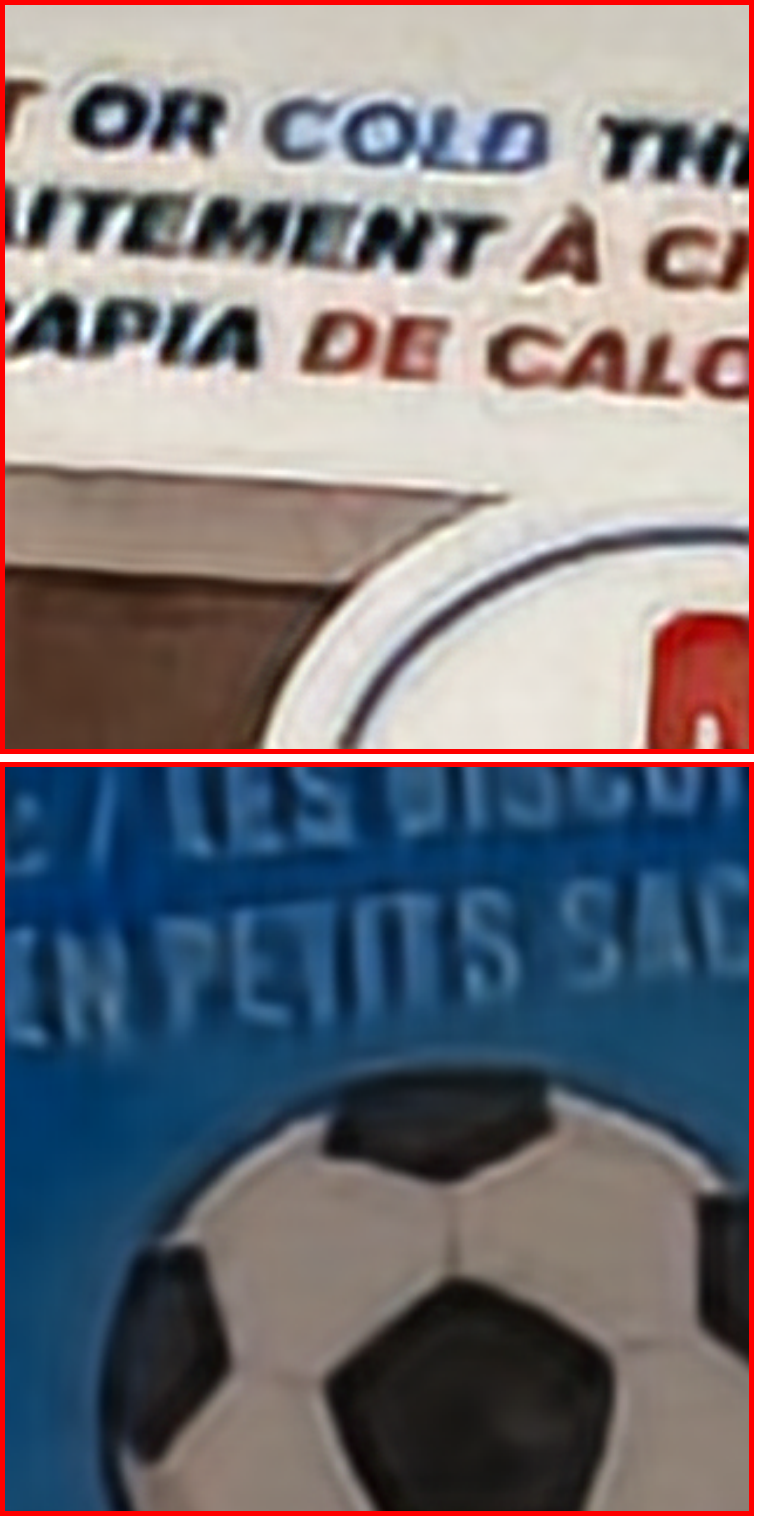}
      \caption*{{RCAN w/ Cai's\cite{cai2019toward}}}
   \end{subfigure}
   \hfill
   \begin{subfigure}[t]{0.195\textwidth}
      \includegraphics[width=1.0\textwidth]{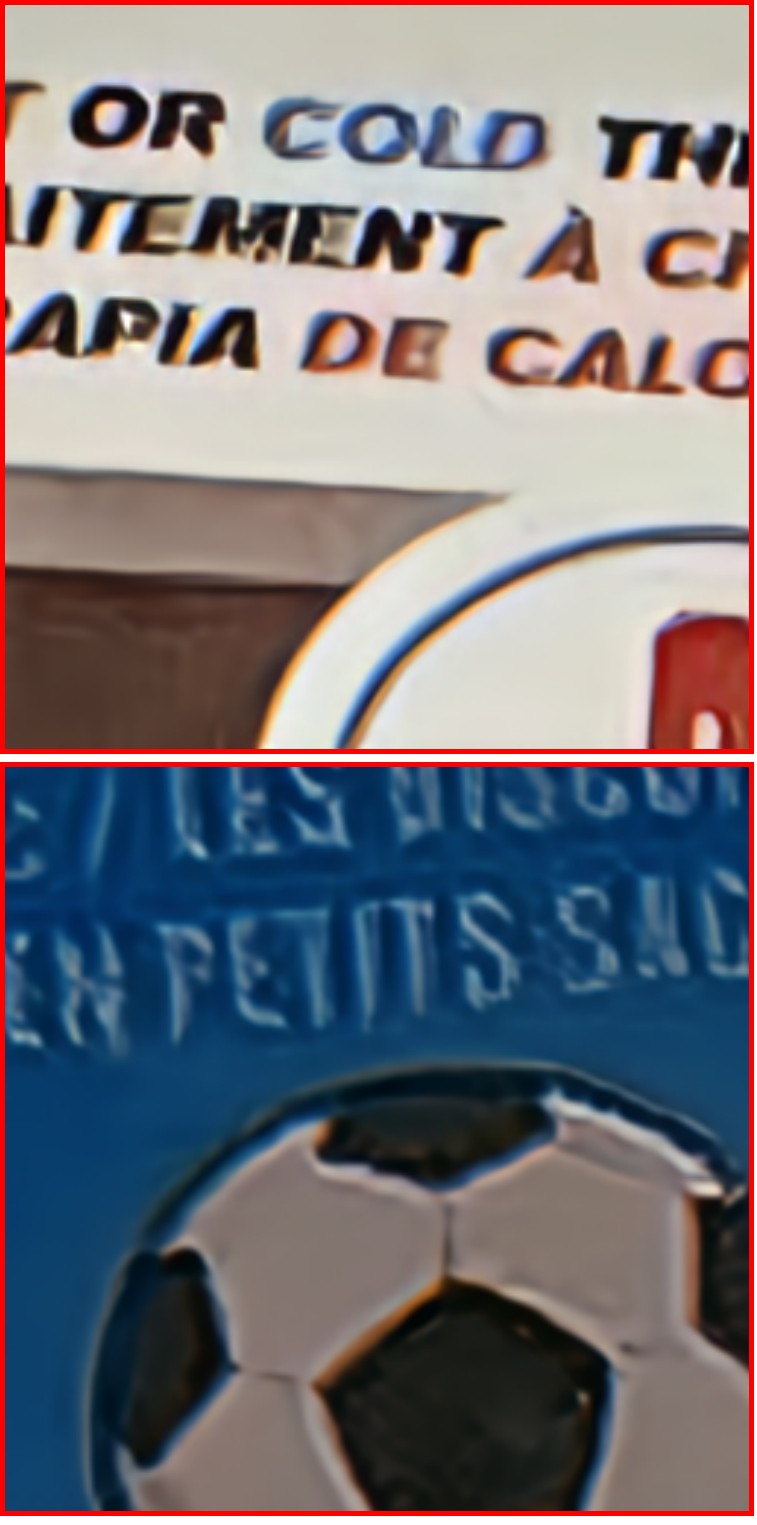}
      \caption*{{ RCAN w/ DRTI}}
   \end{subfigure}

   \caption{SR results of training RCAN by different DSRL camera-captured datasets but inferring on the images captured by iPhone 7plus and iPhone 11.
   Note superior image quality brought by training RCAN with the proposed DRTI datasets over other camera-captured datasets.}

   \label{Figure11_iphone_comparison}
   % \vspace{-0.5cm}
\end{figure*}

\begin{figure*}[t]
   \centering
   \includegraphics[width=1\textwidth]{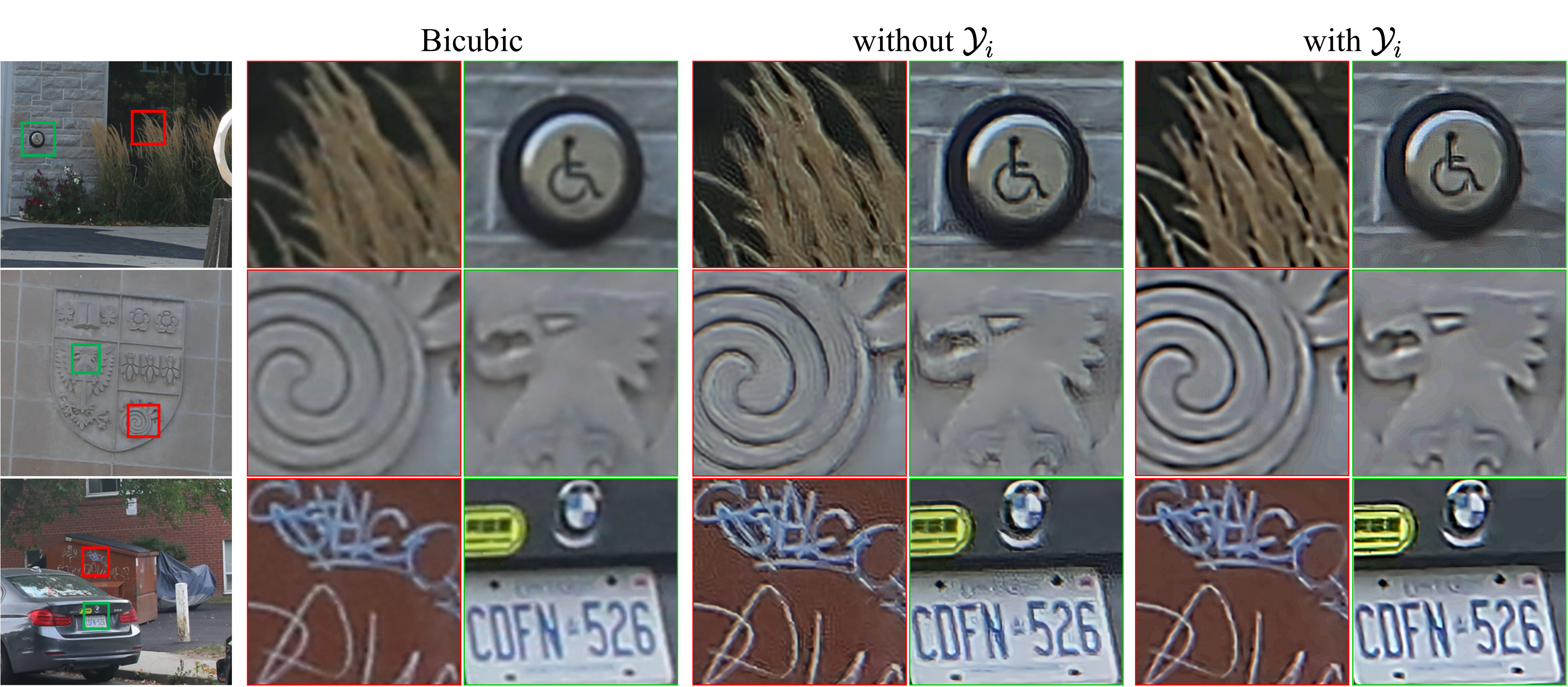}
   \caption{Comparison of $\times 4$ SR results without and with using digital HR images in supervised deep learning.}
   \label{Figure12_ablation}
   % \vspace{-0.5cm}
\end{figure*}

\subsection{Results of superresolving real-world images}
\label{sec_evaluate_real}
%------------------------------------------------------------------------
Our DRTI dataset distinguishes itself from previous camera-acquired training datasets (\eg \cite{zhang2019zoom,cai2019toward})
for the SR task in that it has much higher sub-pixel precision of LR$\sim$HR registration and an extra high qulaity reference for model training.
These advantages contribute to superior performance of DRTI-trained DCNN models when the inference is carried out on screen images in the DRTI dataset.
A tantalizing question is weather the DRTI-trained model can still keep its competitive edge when being applied to real-world images.
After all, images displayed on monitor are not the same as those captured in real-world.
Can the monitor-induced dual-reference data offset the nuance differences between the screen-displayed and real-world images?

To answer the above question, now we compare how well a given DCNN SR model can superresolve real LR images, if trained by the DRTI dataset versus if trained by other competing datasets.
The mainstream synthetic dataset of bicubic downsampling (BD) is used as a baseline.
We choose two DCNN SR models (EDSR and RCAN) in this comparison study.
Since for real LR images there are no ground truth images, we cannot compute PSNR and SSIM.  Instead, let us visually compare, in Fig.~\ref{Figure9_real_comparison},
the output image qualities of EDSR and RCAN after the networks are trained by the DRTI dataset versus trained by the BD dataset.  Experiments are carried out on superresolving real images captured by three different cameras.  For all three cameras, both the DRTI-trained EDSR and the DRTI-trained RCAN reproduce sharper and cleaner edges and details than the models trained on the synthesized dataset, as being evident in Fig.~\ref{Figure9_real_comparison}. This provides empirical evidence for the generalization capability of the DRTI datasets from screen to real-world images.

In the above experiments of Fig.~\ref{Figure9_real_comparison}, the real-world images for inference and training are captured by the same camera.
Next we investigate what if the inference and training images are captured by different cameras.  To this end we add two camera-captured SR training datasets of Chen \etal \cite{chen2019camera} and Cai \etal \cite{cai2019toward}.
In Chen \etal's dataset, similar to our DRTI, the realistic LR$\sim$HR image pairs are collected under a controlled experimental environment. They printed 100 postercards and captured them twice with lenses of different focal lengths.
Their image pairs are aligned by only using SIFT features.
Cai \etal's \cite{cai2019toward} collected LR$\sim$HR image pairs by capturing real-world static scenes.

We train the RCAN model by Chen \etal's dataset (RCAN w/ Chen's), by Cai \etal's dataset (RCAN w/ Cai's), and by our DRTI dataset (RCAN w/ DRTI) collected by the three cameras listed in Table.~\ref{table_camera}.  These three versions of RCAN are applied to superresolve the LR images in Cai \etal's dataset, and the results are presented in Fig.\ref{Figure10_test_cai}.
As can be observed in Fig.\ref{Figure10_test_cai}, the DRTI-trained model super-resolves high frequency features better than the competitors, which is quite remarkable considering that our method is applied to real-world LR images of an unseen camera. Also somewhat surprisingly, the RCAN model trained by Cai \etal's dataset is inferior to our DRTI method in visual quality, even though the images for training and inference are captured by the same camera used by Cai \etal.  This poor SR performance is due to the fact that the captured LR and HR images in \cite{chen2019camera} and \cite{cai2019toward} are not precisely aligned.

In the above experiments the training and inference images are all captured by DSRL cameras.  Now we present, in Fig.~\ref{Figure11_iphone_comparison}, the experimental results for two smartphone cameras (iPhone 7plus and iPhone 11).
The three RCAN models are the same as in Fig.\ref{Figure10_test_cai}, but applied to superresolve images taken by smartphones.
It is evident in Fig.~\ref{Figure11_iphone_comparison} that the DRTI-trained SR model recovers sharper and cleaner details with less artifacts than the other two models, when they are applied to novel cameras whose images are unseen during the training.
All these test images are outside of our DRTI-trained dataset. These experiments offer additional empirical evidence for the robustness of the DRTI training data generation strategy.

\begin{table}[h]
   \centering

   \caption{Average PSNR/SSIM ($\times 4$) for cross-camera evaluation. All models are of RCAN structure.}
   \scalebox{1}{
   \begin{tabular}{||c|c|c|c||}
   \hline\hline
   \bfseries \diagbox{infer}{train} & Sony Nex& Canon 5D  & Sony Alpha\\
   % \cline{2-4}
   \hline
   Sony Nex  &	\textbf{33.15/0.877}  & 32.55/0.809   &   32.78/0.893 \\\hline
   Canon 5D &	31.71/0.862	  & \textbf{33.24/0.853}  & 31.90/0.872   \\\hline
   Sony Alpha	     &	32.14/0.868   & 32.17/0.801  & \textbf{33.21/0.903}   \\\hline
   \hline
   \end{tabular}}
   \label{table_camera_cross}
\end{table}
The above cross-camera SR performance evaluations can only be done visually for the lack of ground truth.  However, as the LR and HR images in our DRTI datasets can be precisely aligned by the novel dual-domain registration algorithm, we can consider the captured HR images as ground truth so that quantitative evaluations become possible for real camera data.  Specifically, we train three RCAN models by three camera-specific DRTI datasets respectively, and then test them on images drawn from all the three DRTI datasets that are unseen in the training.
Table.~\ref{table_camera_cross} reports the PSNR and SSIM values of the test results.

As expected, the SR performance deteriorates when the training and inference images are from two different cameras (see Table.~\ref{table_camera_cross}). This is true if the two cameras are made by the same manufacturer (0.37 dB gap between Sony Nex-6 and Sony Alpha-a7R\uppercase\expandafter{\romannumeral2}).
The performance gap between different camera manufacturers becomes larger (e.g., 0.60 dB gap between Sony Nex-6 and Canon 5D-Mark\uppercase\expandafter{\romannumeral3}).
As illustrated in the last row of the Table.~\ref{table_camera_cross}, both models trained by Sony Nex-6 DRTI dataset and Canon 5D-Mark\uppercase\expandafter{\romannumeral3} DRTI dataset are not nearly as good as the Sony Alpha-a7R\uppercase\expandafter{\romannumeral2} DRTI trained model.

The above observations suggest that for best performance the DCNN SR model trained by images captured by a camera should be applied to images of the same camera.
Thanks to the high data collection throughput of the DRTI acquisition system,
it can be easily employed to collect a large set of LR$\sim$HR image pairs for any given camera. It only takes 4 hours (including the time of the system installation and calibration) to collect 1800 image pairs for three cameras.
Therefore, it is practical to collect training datasets for each of the intended cameras and build a camera-specific DCNN SR model for it.

\subsection{Ablation study of digital files}\label{sec_ablation}

To verify the advantages of the strategy by training SR models with digital HR images, we compare RCAN models trained on the proposed DRTI dataset
with and without the digital HR reference $\mathcal{Y}_i$. As shown in Fig.~\ref{Figure12_ablation}, the results of the model without $\mathcal{Y}_i$ have some objectionable high-frequency artifacts around
the edges. In contrast, the digital HR images trained model can produce clear results meanwhile erasing the unintended high-frequency blemishes.
These improvements demonstrate the effectiveness of the proposed dual references for SR model training. With the dual-reference strategy,
the unintended artifacts of captured image pairs can be suppressed, while the super high quality reference $\mathcal{Y}_i$ can boost the performance of the trained SR models.

\subsection{Limitations}
In pursuing of maximum registration precision of real-world LR and HR image pairs, we choose to shoot a ultra high definition planar monitor in a controlled laboratory setting.
An obvious drawback of this approach is that the captured screen images may differ from those captured in the corresponding real scene.
For instance, the captured screen images could have a narrower dynamic range than their counterparts captured in the wild. 
However, as we explained in the introduction in detail, capturing screen images brings the benefits of more
precisely registered LR and HR image pairs, and the dual-reference strategy.
On balance these benefits overweigh the limitations of screen-captured training images.

%------------------------------------------------------------------------
\section{Conclusion}\label{conclusion}
The proposed DRTI training data generation process for deep learning based super-resolution achieves unprecedented alignment accuracy between LR and HR image pairs captured by cameras.
This unique advantage contributes to superior performance of DRTI-trained DCNN SR models over the same models trained by other LR$\sim$HR datasets.
In addition, it is easy to switch cameras in the DRTI acquisition system and collect a large number of paired LR and HR images for any target camera at low cost and high data throughput.
This offers a practical way of training camera-specific DCNN SR models to eliminate the discrepancies between the training and inference data in pursuit of best possible SR results.

% \section*{Acknowledgments}
% This should be a simple paragraph before the References to thank those individuals and institutions who have supported your work on this article.

% {\appendix[Proof of the Zonklar Equations]
% Use $\backslash${\tt{appendix}} if you have a single appendix:
% Do not use $\backslash${\tt{section}} anymore after $\backslash${\tt{appendix}}, only $\backslash${\tt{section*}}.
% If you have multiple appendixes use $\backslash${\tt{appendices}} then use $\backslash${\tt{section}} to start each appendix.
% You must declare a $\backslash${\tt{section}} before using any $\backslash${\tt{subsection}} or using $\backslash${\tt{label}} ($\backslash${\tt{appendices}} by itself
%  starts a section numbered zero.)}


\bibliographystyle{IEEEtran}
\bibliography{DRTI_TIP_Final}

\begin{thebibliography}{10}
\providecommand{\url}[1]{#1}
\csname url@samestyle\endcsname
\providecommand{\newblock}{\relax}
\providecommand{\bibinfo}[2]{#2}
\providecommand{\BIBentrySTDinterwordspacing}{\spaceskip=0pt\relax}
\providecommand{\BIBentryALTinterwordstretchfactor}{4}
\providecommand{\BIBentryALTinterwordspacing}{\spaceskip=\fontdimen2\font plus
\BIBentryALTinterwordstretchfactor\fontdimen3\font minus
  \fontdimen4\font\relax}
\providecommand{\BIBforeignlanguage}[2]{{%
\expandafter\ifx\csname l@#1\endcsname\relax
\typeout{** WARNING: IEEEtran.bst: No hyphenation pattern has been}%
\typeout{** loaded for the language `#1'. Using the pattern for}%
\typeout{** the default language instead.}%
\else
\language=\csname l@#1\endcsname
\fi
#2}}
\providecommand{\BIBdecl}{\relax}
\BIBdecl

\bibitem{huang2015single}
J.-B. Huang, A.~Singh, and N.~Ahuja, ``Single image super-resolution from
  transformed self-exemplars,'' in \emph{Proceedings of the IEEE conference on
  computer vision and pattern recognition}, 2015, pp. 5197--5206.

\bibitem{yang2010image}
J.~Yang, J.~Wright, T.~S. Huang, and Y.~Ma, ``Image super-resolution via sparse
  representation,'' \emph{IEEE transactions on image processing}, vol.~19,
  no.~11, pp. 2861--2873, 2010.

\bibitem{dong2015PAMI}
C.~Dong, C.~C. Loy, K.~He, and X.~Tang, ``Image super-resolution using deep
  convolutional networks,'' \emph{IEEE transactions on pattern analysis and
  machine intelligence}, vol.~38, no.~2, pp. 295--307, 2015.

\bibitem{ledig2017photo}
C.~Ledig, L.~Theis, F.~Husz{\'a}r, J.~Caballero, A.~Cunningham, A.~Acosta,
  A.~Aitken, A.~Tejani, J.~Totz, Z.~Wang \emph{et~al.}, ``Photo-realistic
  single image super-resolution using a generative adversarial network,'' in
  \emph{Proceedings of the IEEE conference on computer vision and pattern
  recognition}, 2017, pp. 4681--4690.

\bibitem{zhang2018ECCV}
Y.~Zhang, K.~Li, K.~Li, L.~Wang, B.~Zhong, and Y.~Fu, ``Image super-resolution
  using very deep residual channel attention networks,'' in \emph{Proceedings
  of the European Conference on Computer Vision (ECCV)}, 2018, pp. 286--301.

\bibitem{lim2017enhanced}
B.~Lim, S.~Son, H.~Kim, S.~Nah, and K.~Mu~Lee, ``Enhanced deep residual
  networks for single image super-resolution,'' in \emph{Proceedings of the
  IEEE conference on computer vision and pattern recognition workshops}, 2017,
  pp. 136--144.

\bibitem{zhang2018image}
Y.~Zhang, K.~Li, K.~Li, L.~Wang, B.~Zhong, and Y.~Fu, ``Image super-resolution
  using very deep residual channel attention networks,'' in \emph{Proceedings
  of the European Conference on Computer Vision (ECCV)}, 2018, pp. 286--301.

\bibitem{yang2020learning}
F.~Yang, H.~Yang, J.~Fu, H.~Lu, and B.~Guo, ``Learning texture transformer
  network for image super-resolution,'' in \emph{Proceedings of the IEEE/CVF
  Conference on Computer Vision and Pattern Recognition}, 2020, pp. 5791--5800.

\bibitem{kim2016accurate}
J.~Kim, J.~Kwon~Lee, and K.~Mu~Lee, ``Accurate image super-resolution using
  very deep convolutional networks,'' in \emph{Proceedings of the IEEE
  conference on computer vision and pattern recognition}, 2016, pp. 1646--1654.

\bibitem{johnson2016perceptual}
J.~Johnson, A.~Alahi, and L.~Fei-Fei, ``Perceptual losses for real-time style
  transfer and super-resolution,'' in \emph{European conference on computer
  vision}.\hskip 1em plus 0.5em minus 0.4em\relax Springer, 2016, pp. 694--711.

\bibitem{sajjadi2017enhancenet}
M.~S. Sajjadi, B.~Scholkopf, and M.~Hirsch, ``Enhancenet: Single image
  super-resolution through automated texture synthesis,'' in \emph{Proceedings
  of the IEEE International Conference on Computer Vision}, 2017, pp.
  4491--4500.

\bibitem{Wang_2018_ECCV_Workshops}
X.~Wang, K.~Yu, S.~Wu, J.~Gu, Y.~Liu, C.~Dong, Y.~Qiao, and C.~Change~Loy,
  ``Esrgan: Enhanced super-resolution generative adversarial networks,'' in
  \emph{Proceedings of the European Conference on Computer Vision (ECCV)
  Workshops}, September 2018.

\bibitem{chen2019camera}
C.~Chen, Z.~Xiong, X.~Tian, Z.-J. Zha, and F.~Wu, ``Camera lens
  super-resolution,'' 2019.

\bibitem{cai2019toward}
J.~Cai, H.~Zeng, H.~Yong, Z.~Cao, and L.~Zhang, ``Toward real-world single
  image super-resolution: A new benchmark and a new model,'' in
  \emph{Proceedings of the IEEE International Conference on Computer Vision},
  2019.

\bibitem{zhang2019zoom}
X.~C. Zhang, Q.~Chen, R.~Ng, and V.~Koltun, ``Zoom to learn, learn to zoom,''
  2019.

\bibitem{zhangwu2008}
X.~Zhang and X.~Wu, ``Image interpolation by adaptive 2-d autoregressive
  modeling and soft-decision estimation,'' \emph{IEEE Transactions on Image
  Processing}, vol.~17, no.~6, pp. 887--896, 2008.

\bibitem{sun2008image}
J.~Sun, Z.~Xu, and H.-Y. Shum, ``Image super-resolution using gradient profile
  prior,'' in \emph{2008 IEEE Conference on Computer Vision and Pattern
  Recognition}.\hskip 1em plus 0.5em minus 0.4em\relax IEEE, 2008, pp. 1--8.

\bibitem{liu2018non}
D.~Liu, B.~Wen, Y.~Fan, C.~C. Loy, and T.~S. Huang, ``Non-local recurrent
  network for image restoration,'' \emph{Advances in neural information
  processing systems}, vol.~31, 2018.

\bibitem{mei2021image}
Y.~Mei, Y.~Fan, and Y.~Zhou, ``Image super-resolution with non-local sparse
  attention,'' in \emph{Proceedings of the IEEE/CVF Conference on Computer
  Vision and Pattern Recognition}, 2021, pp. 3517--3526.

\bibitem{zhou2020cross}
S.~Zhou, J.~Zhang, W.~Zuo, and C.~C. Loy, ``Cross-scale internal graph neural
  network for image super-resolution,'' \emph{Advances in neural information
  processing systems}, vol.~33, pp. 3499--3509, 2020.

\bibitem{dosovitskiy2020vit}
A.~Dosovitskiy, L.~Beyer, A.~Kolesnikov, D.~Weissenborn, X.~Zhai,
  T.~Unterthiner, M.~Dehghani, M.~Minderer, G.~Heigold, S.~Gelly, J.~Uszkoreit,
  and N.~Houlsby, ``An image is worth 16x16 words: Transformers for image
  recognition at scale,'' \emph{ICLR}, 2021.

\bibitem{liang2021swinir}
J.~Liang, J.~Cao, G.~Sun, K.~Zhang, L.~Van~Gool, and R.~Timofte, ``Swinir:
  Image restoration using swin transformer,'' in \emph{IEEE International
  Conference on Computer Vision Workshops}, 2021.

\bibitem{zamir2021restormer}
S.~W. Zamir, A.~Arora, S.~Khan, M.~Hayat, F.~S. Khan, and M.-H. Yang,
  ``Restormer: Efficient transformer for high-resolution image restoration,''
  \emph{arXiv preprint arXiv:2111.09881}, 2021.

\bibitem{Ulyanov_2018_CVPR}
D.~Ulyanov, A.~Vedaldi, and V.~Lempitsky, ``Deep image prior,'' in
  \emph{Proceedings of the IEEE Conference on Computer Vision and Pattern
  Recognition (CVPR)}, June 2018.

\bibitem{Shocher_2018_CVPR}
A.~Shocher, N.~Cohen, and M.~Irani, ``“zero-shot” super-resolution using
  deep internal learning,'' in \emph{Proceedings of the IEEE Conference on
  Computer Vision and Pattern Recognition (CVPR)}, June 2018.

\bibitem{Soh_2020_CVPR}
J.~W. Soh, S.~Cho, and N.~I. Cho, ``Meta-transfer learning for zero-shot
  super-resolution,'' in \emph{Proceedings of the IEEE/CVF Conference on
  Computer Vision and Pattern Recognition (CVPR)}, June 2020.

\bibitem{bevilacqua2012low}
M.~Bevilacqua, A.~Roumy, C.~Guillemot, and M.~L. Alberi-Morel, ``Low-complexity
  single-image super-resolution based on nonnegative neighbor embedding,''
  2012.

\bibitem{zeyde2010single}
E.~M. Zeyde~Roman and P.~Matan, ``On single image scale-up using
  sparse-representations,'' in \emph{International conference on curves and
  surfaces}.\hskip 1em plus 0.5em minus 0.4em\relax Springer, 2010, pp.
  711--730.

\bibitem{timofte2017ntire}
R.~Timofte, E.~Agustsson, L.~Van~Gool, M.-H. Yang, and L.~Zhang, ``Ntire 2017
  challenge on single image super-resolution: Methods and results,'' in
  \emph{Proceedings of the IEEE conference on computer vision and pattern
  recognition workshops}, 2017, pp. 114--125.

\bibitem{dong2012nonlocally}
W.~Dong, L.~Zhang, G.~Shi, and X.~Li, ``Nonlocally centralized sparse
  representation for image restoration,'' \emph{IEEE transactions on Image
  Processing}, vol.~22, no.~4, pp. 1620--1630, 2012.

\bibitem{Xu2019TowardsRS}
X.~Xu, Y.~Ma, and W.~Sun, ``Towards real scene super-resolution with raw
  images,'' \emph{2019 IEEE/CVF Conference on Computer Vision and Pattern
  Recognition (CVPR)}, pp. 1723--1731, 2019.

\bibitem{Jaejun2020Rethink}
J.~Yoo, N.~Ahn, and K.-A. Sohn, ``Rethinking data augmentation for image
  super-resolution: A comprehensive analysis and a new strategy,'' 04 2020.

\bibitem{jeon2018enhancing}
D.~S. Jeon, S.-H. Baek, I.~Choi, and M.~H. Kim, ``Enhancing the spatial
  resolution of stereo images using a parallax prior,'' in \emph{Proceedings of
  the IEEE Conference on Computer Vision and Pattern Recognition}, 2018, pp.
  1721--1730.

\bibitem{wang2019learning}
L.~Wang, Y.~Wang, Z.~Liang, Z.~Lin, J.~Yang, W.~An, and Y.~Guo, ``Learning
  parallax attention for stereo image super-resolution,'' in \emph{Proceedings
  of the IEEE Conference on Computer Vision and Pattern Recognition}, 2019, pp.
  12\,250--12\,259.

\bibitem{Manuel2019}
M.~Fritsche, S.~Gu, and R.~Timofte, ``Frequency separation for real-world
  super-resolution,'' 10 2019, pp. 3599--3608.

\bibitem{yuan2018unsupervised}
Y.~Yuan, S.~Liu, J.~Zhang, Y.~Zhang, C.~Dong, and L.~Lin, ``Unsupervised image
  super-resolution using cycle-in-cycle generative adversarial networks,'' in
  \emph{Proceedings of the IEEE Conference on Computer Vision and Pattern
  Recognition Workshops}, 2018, pp. 701--710.

\bibitem{Adrian2018}
A.~Bulat, J.~Yang, and G.~Tzimiropoulos, ``To learn image super-resolution, use
  a gan to learn how to do image degradation first,'' in \emph{Computer Vision
  -- ECCV 2018}.\hskip 1em plus 0.5em minus 0.4em\relax Cham: Springer
  International Publishing, 2018, pp. 187--202.

\bibitem{Qu2016}
C.~{Qu}, D.~{Luo}, E.~{Monari}, T.~{Schuchert}, and J.~{Beyerer}, ``Capturing
  ground truth super-resolution data,'' in \emph{2016 IEEE International
  Conference on Image Processing (ICIP)}, 2016, pp. 2812--2816.

\bibitem{Thomas2020}
T.~{Köhler}, M.~{Bätz}, F.~{Naderi}, A.~{Kaup}, A.~{Maier}, and C.~{Riess},
  ``Toward bridging the simulated-to-real gap: Benchmarking super-resolution on
  real data,'' \emph{IEEE Transactions on Pattern Analysis and Machine
  Intelligence}, vol.~42, no.~11, pp. 2944--2959, 2020.

\bibitem{lowe2004distinctive}
D.~G. Lowe, ``Distinctive image features from scale-invariant keypoints,''
  \emph{International journal of computer vision}, vol.~60, no.~2, pp. 91--110,
  2004.

\bibitem{zhang2000flexible}
Z.~Zhang, ``A flexible new technique for camera calibration,'' \emph{IEEE
  Transactions on pattern analysis and machine intelligence}, vol.~22, no.~11,
  pp. 1330--1334, 2000.

\bibitem{bay2006surf}
H.~Bay, T.~Tuytelaars, and L.~Van~Gool, ``Surf: Speeded up robust features,''
  in \emph{European conference on computer vision}.\hskip 1em plus 0.5em minus
  0.4em\relax Springer, 2006, pp. 404--417.

\bibitem{Bracewell1993AffineTF}
R.~N. Bracewell, K.-Y. Chang, A.~K. Jha, and Y.-H. Wang, ``Affine theorem for
  two-dimensional fourier transform,'' \emph{Electronics Letters}, vol.~29, pp.
  304--304, 1993.

\bibitem{Mehdi2021}
S.~M. Ayyoubzadeh and X.~Wu, ``High frequency detail accentuation in cnn image
  restoration,'' \emph{IEEE Transactions on Image Processing}, vol.~30, pp.
  8836--8846, 2021.

\bibitem{upsample_sa2008}
Q.~Shan, Z.~Li, J.~Jia, and C.-K. Tang, ``Fast image/video upsampling,''
  \emph{ACM Transactions on Graphics (SIGGRAPH ASIA)}, 2008.

\bibitem{yu2012robust}
L.~Yu, H.~Xu, Y.~Xu, and X.~Yang, ``Robust single image super-resolution based
  on gradient enhancement,'' in \emph{Proceedings of The 2012 Asia Pacific
  Signal and Information Processing Association Annual Summit and
  Conference}.\hskip 1em plus 0.5em minus 0.4em\relax IEEE, 2012, pp. 1--6.

\bibitem{zhang2021context}
Y.~Zhang, D.~Wei, C.~Qin, H.~Wang, H.~Pfister, and Y.~Fu, ``Context reasoning
  attention network for image super-resolution,'' in \emph{Proceedings of the
  IEEE/CVF International Conference on Computer Vision}, 2021, pp. 4278--4287.

\bibitem{kingma2014adam}
D.~P. Kingma and J.~Ba, ``Adam: A method for stochastic optimization,''
  \emph{arXiv preprint arXiv:1412.6980}, 2014.

\end{thebibliography}

\end{document}